\documentclass[preprint,pra,showkeys,nofootinbib,superscriptaddress]{revtex4}
\usepackage[english]{babel}
\usepackage[utf8]{inputenc}
\usepackage{graphicx}
\usepackage{comment}
\usepackage{multirow}

\input{Qcircuit} 

\begin{document}

\title{Quantum computing applied to calculations of molecular energies: CH$_{2}$ benchmark}
\author{Libor Veis}
\email{libor.veis@jh-inst.cas.cz}
\affiliation{Charles University in Prague, Faculty of Science, Department of Physical and Macromolecular Chemistry, Hlavova 8, 12840 Prague 2, Czech Republic}
\affiliation{J. Heyrovsk\'{y} Institute of Physical Chemistry, Academy of Sciences of the Czech \mbox{Republic, v.v.i.}, Dolej\v{s}kova 3, 18223 Prague 8, Czech Republic}

\author{Ji\v{r}\'{i} Pittner}
\email{jiri.pittner@jh-inst.cas.cz, corresponding author}
\affiliation{J. Heyrovsk\'{y} Institute of Physical Chemistry, Academy of Sciences of the Czech \mbox{Republic, v.v.i.}, Dolej\v{s}kova 3, 18223 Prague 8, Czech Republic}

\date{\today}

\begin{abstract}
Quantum computers are appealing for their ability to solve some tasks much faster than their classical counterparts. It was shown in [Aspuru-Guzik et al., \textit{Science} \textbf{309}, 1704 (2005)] that they, if available, would be able to perform the full configuration interaction (FCI) energy calculations with a polynomial scaling. This is in contrast to conventional computers where FCI scales exponentially. We have developed a code for simulation of quantum computers and implemented our version of the quantum full configuration interaction algorithm. We provide a detailed description of this algorithm and the results of the assessment of its performance on the four lowest lying electronic states of CH$_{2}$ molecule. This molecule was chosen as a benchmark, since its two lowest lying $^{1}A_{1}$ states exhibit a multireference character at the equilibrium geometry. It has been shown that with a suitably chosen initial state of the quantum register, one is able to achieve 
the probability amplification regime of the iterative phase estimation algorithm even in this case.
\end{abstract}


\maketitle

\section{Introduction}
Quantum chemical computations on conventional computers achieved a huge success in the last decades, both due to the enormous progress in computer technology and in the development of approximate quantum chemical methods. They became an indispensable part of basic as well as applied chemical research. The situation is however far from ideal. Exact solution of the Schr\"{o}dinger equation within a finite one-particle basis set (full configuration interaction, FCI) is limited only to the smallest systems (diatomics, triatomics) because of the exponential scaling of its computational cost. Approximate methods with polynomial scaling must be used instead, of course at the cost of accuracy and reliability.

Quantum computers on the other hand, if available, would offer exponential speedup for certain types of problems \cite{nielsen_chuang}. The most prominent example is the famous Shor's algorithm \cite{shor_1994, shor_1997} for factoring integers, with potentially far-reaching consequences for cryptography. Another promising application of quantum computers, firstly suggested by Feynman \cite{Feynman_1982}, is \textit{efficient} simulation of another quantum systems \cite{lloyd_1996, zalka_1998, ortiz_2001, somma_2002, abrams_1997}. Abrams's and Lloyd's phase estimation method \cite{abrams_1999} represents the polynomially scaling quantum algorithm for obtaining eigenvalues of local Hamiltonians.

The first work connecting quantum computation and quantum chemistry by Lidar and Wang \cite{lidar_1999} concerned the \textit{efficient} calculations of thermal rate constants of chemical reactions. This work in fact founded the new field of quantum chemistry: ``Quantum chemistry on quantum computers". Six years later, Aspuru-Guzik et al. \cite{aspuru-guzik_2005} applied the Abrams's and Lloyd's algorithm \cite{abrams_1999} to quantum chemical energy calculations and proposed that quantum computers with 30 to 100 (noise free) qubits could already exceed the limits of classical FCI calculations. Since these two pioneering works, other papers involving energy calculations of excited states \cite{wang_2008}, quantum chemical dynamics \cite{kassal_2008}, calculations of molecular properties \cite{kassal_2009}, state preparation \cite{ward_2009, wang_2009} or global minima search \cite{zhu_2009} were published. Whitfield et al. \cite{whitfield_2010} provided the detailed description of the quantum FCI algorithm \cite{aspuru-guzik_2005} with emphasis on the factorization of an exponential of a Hamiltonian operator to elementary one and two qubit gates. Analysis of the resource requirements for the ground state energy calculations of the one dimensional transverse Ising model which took into account also the error correction was done by Clark et al. \cite{clark_2009}. An up to date summary of quantum chemical algorithms for quantum computers could be found in a recent review \cite{kassal_review}. 

Very recently, also the first proof-of-principle experiments representing quantum chemical calculations on quantum computers have appeared \cite{lanyon_2010, du_2010}. These are energy calculations of hydrogen molecule in a minimal basis set using optical quantum computer \cite{lanyon_2010} and NMR quantum computer \cite{du_2010}. In spite of the fact that these experiments treated only the simplest case of hydrogen molecule and did not have the resources to implement quantum error correction, it seems that simulations of physical phenomena on quantum computers might become available in foreseeable future \cite{nori_science_2009}.

The paper is organized as follows. The second section is devoted to the overview of quantum full configuration interaction algorithm, with the emphasis on the iterative phase estimation, since we feel that a more detailed description of the algorithm would be useful. The third section involves the application of our version of this algorithm to the four lowest lying electronic states of CH$_{2}$ molecule.

\section{Overview of quantum full configuration interaction (QFCI) algorithm}
The first question to answer, when thinking about quantum chemical calculations on quantum computers, is how to map the quantum chemical wave function onto the register of quantum bits (qubits). Several different mappings have been proposed for this purpose \cite{aspuru-guzik_2005,wang_2008}. Throughout this paper, we work with the simplest in terms of factorization of an exponential of the Hamiltonian to elementary gates and therefore  the simplest for experimental realization, but the least economical one, so called direct mapping. In this approach, individual spin orbitals are directly assigned to qubits, since each spin orbital can be either occupied or unoccupied, corresponding to $|1\rangle$ or $|0\rangle$ states. Non-economical property lies in the fact that it actually maps the whole Fock space of the system (states with different number of electrons) on the Hilbert space of the quantum computer. A compact mapping from a subspace of fixed-electron-number wave functions, spin-adapted or even symmetry-adapted wave functions to the qubits have been also proposed \cite{aspuru-guzik_2005,wang_2008}, but the general factorization schemes for these mappings have not been discovered yet. 

Overview of the QFCI algorithm is divided into three subsections. In the first, state preparation is briefly discussed, the second is devoted to the phase estimation procedure and its improved iterative variant, which is the heart of the algorithm. The last section shortly presents factorization of the exponential of the molecular electronic Hamiltonian to elementary gates.

\subsection{Initial state preparation}
Quantum FCI algorithm must be started with some initial guess state. Generally it holds that the closer is the initial guess to the exact wave function corresponding to calculated energy, the higher is the success probability of observing the energy. As is confirmed further in the paper, the Hartree-Fock guess, which is the first thing one tries to use, may not be successful especially in situations, where correlation (particularly the static one) plays an important role. In these situations, initial guess states from more sophisticated \textit{polynomially} scaling methods can be used (e.g. CASSCF in a limited orbital space). 

Preparing a general initial state (vector from the Hilbert space of $n$ qubits) is in fact a difficult task as this vector can contain up to $2^{n}$ non-zero components. Fortunately, initial guesses which include only a few determinants in a superposition (about 10 in our most difficult case) are sufficient for most purposes of quantum chemistry. These states can be prepared e.g. with the procedure described by Ortiz et al. \cite{ortiz_2001} which scales as $\mathcal{O}(N^{2})$ in the number of determinants $N$. Preparation of general molecular-like states from the combinatorial space of dimension $\big(\begin{array}{c} n \\ m \end{array}\big)$ corresponding to distributing $m$ electrons among $n$ spin orbitals was presented in \cite{wang_2009}. Preparation of many-particle states in a superposition on a lattice which can be then propagated by quantum chemical dynamics algorithm \cite{kassal_2008} was studied in \cite{ward_2009}.

Completely different approach is based on the adiabatic quantum computation \cite{fahri_2000,aspuru-guzik_2005}, which is an alternative to the classical gate model. Here one starts with a Hamiltonian, whose ground state is easy to construct, and gradually varies this Hamiltonian into the final exact one, where its ground state encodes the solution of the computational problem. If the variation of the Hamiltonian is slow enough, the system will remain in the ground state according to the adiabatic theorem. It has been shown that any quantum circuit can be simulated by an adiabatic quantum computation with an appropriately constructed Hamiltonian \cite{Aharonov,Kitaev}. When preparing initial guess states for the quantum FCI algorithm, one starts with the Hartree-Fock Hamiltonian and evolves to the exact one \cite{aspuru-guzik_2005}. This approach seems to be very promising especially due to the possible experimentally more accessible realization. However, its scaling is rather unclear, because the maximum allowed speed of the adiabatic evolution depends on the minimum of the energy difference between the ground and the first excited state along the adiabatic path.  

\subsection{Phase estimation}
The phase estimation algorithm is an efficient quantum algorithm for obtaining the eigenvalue of an unitary operator $\hat{U}$, based on a given initial guess of the corresponding eigenvector. It is actually also the key part of the famous Shor's factoring algorithm \cite{shor_1994, shor_1997}. For a comprehensive description of the phase estimation procedure and citations to original papers see Nielsen and Chuang \cite{nielsen_chuang}. 

Suppose that $| u \rangle$ is an eigenvector of $\hat{U}$ and that it holds

\begin{equation}
 \hat{U} | u \rangle = e^{2\pi i \phi} | u \rangle, \qquad \phi \in \langle 0, 1),
\end{equation}

\noindent
where $\phi$ is the phase which is estimated by the algorithm. Quantum register is divided into two parts. The first part is the read-out part composed of $m$ qubits on which the binary representation of $\phi$ is measured at the end and which is initialized to the state $| 0 \rangle^{\otimes m}$. The second part contains at the beginning of the procedure the corresponding eigenvector $| u \rangle$. The initial state of the quantum register thus reads

\begin{equation}
 | \mathrm{reg} \rangle = \underbrace{| 0 \rangle \otimes | 0 \rangle \otimes \ldots \otimes | 0 \rangle}_{m~\rm{qubits}} \otimes | u \rangle, \quad \mathrm{shortly} \quad | 0 \rangle | u \rangle  
\end{equation}

\noindent
Application of Hadamard gates ($\pi/2$ rotations) on all qubits in the first part of the register gives

\begin{equation}
 | \mathrm{reg} \rangle = \frac{1}{\sqrt{2^{m}}} \sum_{j=0}^{2^{m}-1} | j \rangle | u \rangle .
 \label{after_h}
\end{equation}

\noindent
After application of sequence of controlled-$\hat{U}^{2^{k-1}}$ operations ($k$s are nonnegative integers from 1 to $m$), the register is transformed into

\begin{equation}
| \mathrm{reg} \rangle = \frac{1}{\sqrt{2^{m}}} \sum_{j=0}^{2^{m}-1} \hat{U}^{j} | j \rangle | u \rangle = \frac{1}{\sqrt{2^{m}}} \sum_{j=0}^{2^{m}-1} e^{2\pi ij \phi} | j \rangle | u \rangle. 
 \label{after_cu} 
\end{equation}

The heart of the phase estimation algorithm is the inverse quantum Fourier transform (iQFT) \cite{nielsen_chuang} performed on the read-out part of the register which is transformed to $| 2^{m}\phi \rangle | u \rangle$. 

If the phase could be expressed exactly in $m$ bits

\begin{eqnarray}
 \phi & = & 0.\phi_{1}\phi_{2} \ldots \phi_{m} \\
  & = & \frac{\phi_{1}}{2} + \frac{\phi_{2}}{2^{2}} + \ldots + \frac{\phi_{m}}{2^{m}}, \qquad \phi_{i} \in \{0,1\}, 
\end{eqnarray}

\noindent
it (and consequently the eigenvalue) could be recovered with unity probability by a measurement on the first part of the quantum register.

If the desired eigenvector is not known explicitly (as is typically the case in quantum chemistry), we can start the algorithm with an arbitrary vector $| \psi \rangle$, which can be expanded in terms of eigenvectors of $\hat{U}$

\begin{equation}
 | \psi \rangle = \sum_{i} c_{i} | u_{i} \rangle. 
\end{equation}

\noindent
The probability of obtaining $\phi_{i}$ is due to linearity of the algorithm $|c_{i}|^{2}$. It is important to note that the initial guess does not influence the accuracy of the phase, only the probability with which the phase of a particular eigenstate is measured.

The situation is more complicated when $\phi$ cannot be expressed exactly in $m$ bits. Then

\begin{equation}
 \phi = \tilde{\phi} + \delta 2^{-m},  
 \label{reminder}
\end{equation}

\noindent
where $\tilde{\phi} = \phi_{1}\phi_{2}\ldots\phi_{m}$ denotes the first $m$ bits of the binary expansion and $0 \le \delta < 1$ is a remainder. It can be shown (e.g. \cite{dobsicek_phd}) that the sum of success probabilities corresponding to $\phi$ rounded down ($P_{\rm{down}} \Rightarrow \phi_{\rm{measured}} = \tilde{\phi}$) and rounded up ($P_{\rm{up}} \Rightarrow \phi_{\rm{measured}} = \tilde{\phi} + 2^{-m}$) decreases monotonically for increasing $m$ and in the limit $m \rightarrow \infty$ the lower bound is 

\begin{equation}
 P_{\rm{down}}(\delta = 1/2) + P_{\rm{up}}(\delta = 1/2) = \frac{4}{\pi^{2}} + \frac{4}{\pi^{2}} > 0.81.
 \label{sc_factor}
\end{equation} 

Note that linearity of the aforementioned scheme leads to the lower bound for $P_{\mathrm{down}} + P_{\mathrm{up}}$ corresponding to $\phi_{i}$ being $0.81\cdot|c_{i}|^{2}$.

\subsubsection*{Iterative phase estimation (IPEA)}
With the use of semiclassical QFT \cite{griffiths_1996}, the circuit representing the phase estimation algorithm can be greatly simplified, having only one ancillary qubit in the read-out part of the quantum register. The algorithm then proceeds in an iterative manner. The $k$-th iteration of this scheme is presented in Figure \ref{ipea_iteration}. 

The algorithm is iterated backwards from the least significant bits of $\phi$ to the most significant ones, for $k$ going from $m$ to 1. The iteration again starts with the Hadamard gate on the read-out qubit followed by controlled-$\hat{U}^{2^{k-1}}$ operation. The equivalent of QFT is a single qubit $z$-rotation $R_{z}$, whose angle $\omega_{k}$ depends on the results of the previously measured bits

\begin{eqnarray}
 R_{z}(\omega_{k}) & = & \left( \begin{array}{cc} 
                            1 & 0 \\
                            0 & e^{2 \pi i \omega_{k}} \\
                            \end{array} \right) \\              
 \omega_{k} & = & -\sum_{i=2}^{m-k+1} \frac{\phi_{k+i-1}}{2^{i}},
\end{eqnarray} 

\noindent
followed by a Hadamard gate. 

IPEA is in fact completely equivalent to the original (multiqubit) phase estimation \cite{dobsicek_phd}. It thus suffers from the same decreasing of success probability when the phase cannot be expressed exactly in a particular number of bits. One possibility of a success probability amplification is performing more iteration steps (more than is the desired accuracy of $\phi$): when extracting $m' = m + \log (2 + 1/2 \epsilon)$ bits, the phase is accurate to $m$ binary digits with probability at least $1 - \epsilon$ \cite{nielsen_chuang}. This method is however not very useful, since implementing the $\hat{U}^{2^{k-1}}$ gate for large $k$ is the algorithm's bottleneck in a realistic noisy environment \cite{dobsicek_2007}.  

Dob\v{s}\'{i}\v{c}ek et al. \cite{dobsicek_2007} came up with a different approach. They repeated the measurement for the least important bits of the phase binary expansion. Using the majority voting (for bit value 0 or 1), the effective error probability decreases exponentially with the number of repetitions according to the binomial distribution. This measurement repetition only for the few least important bits of $\phi$ is unfortunately possible only if the exact eigenstates of $\hat{U}$ are available.

\begin{figure}[!h]
 \begin{center}
   \includegraphics[width=12cm]{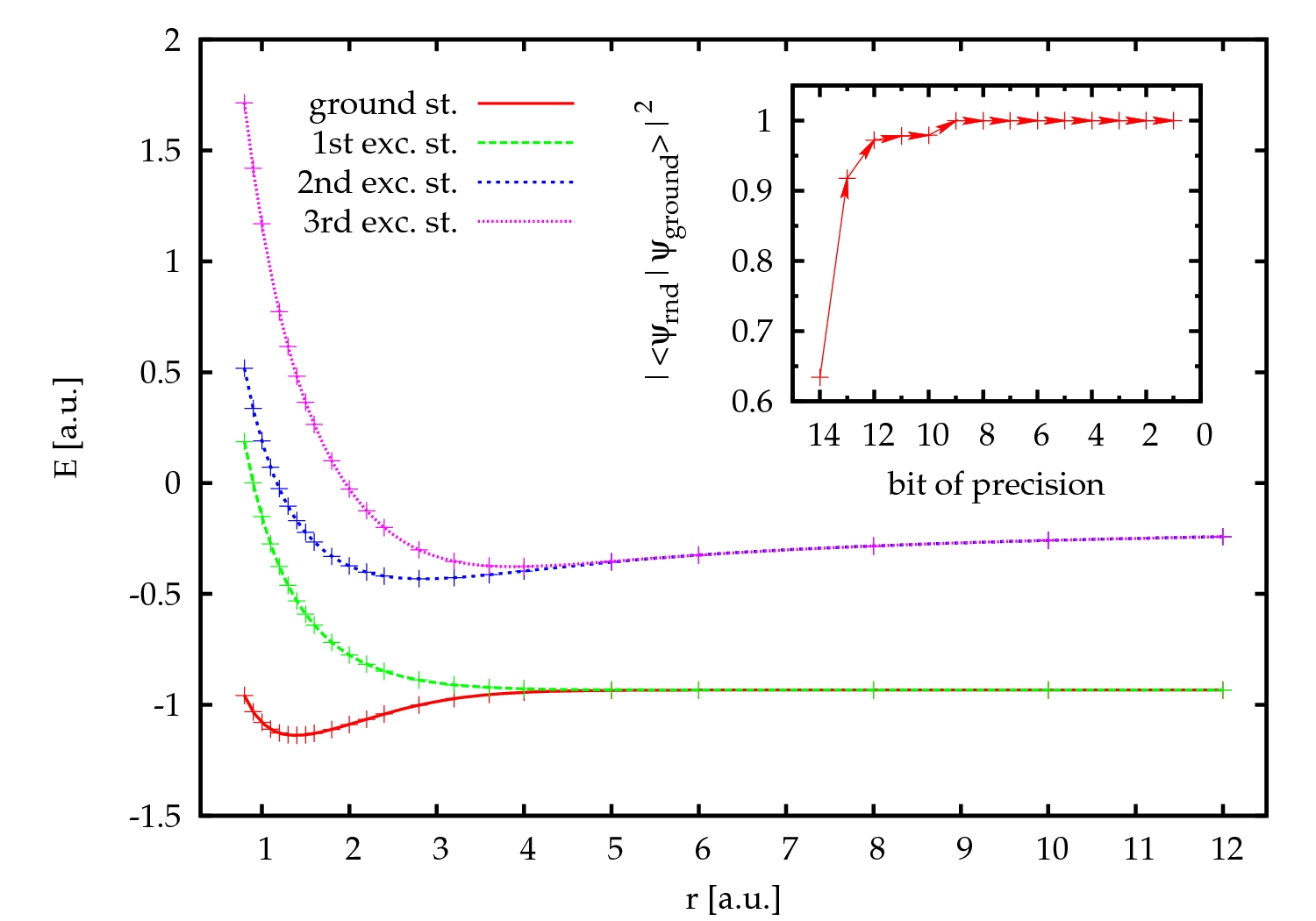}
   \caption{Energies of the four electronic states of H$_{2}$ in STO-3G basis which were obtained by QFCI (IPEA version \textbf{A}) with randomly generated initial guess states. Small figure inside presents the increasing overlap between the actual state of the second part of the quantum register and the exact wave function for one of random runs of the algorithm leading to the ground state.}
   \label{h2_rnd}
 \end{center} 
\end{figure}

When working with general initial states (the case of quantum chemistry), two scenarios presented in Figure \ref{ipea_comparison} come into question. Maintaining the second part of the quantum register during all iterations and amplification of the success probabilities by repeating the whole process is the first possibility. We denote this version as \textbf{A} version of IPEA. The biggest advantage of this approach is that one \textit{always} ends up with one of the eigenstates of $\hat{U}$ in the second part of the quantum register as was the case in the original PEA. This happens through successive collapses of the system state into the corresponding eigensubspace. We demonstrate this approach in Figure \ref{h2_rnd} on the hydrogen molecule in a minimal basis set and \textit{random} initial states (to avoid collapsing into the eigenstate which does not fit into the correct number of electrons - 2, we generated random initial states mixing only the components corresponding to correct number of electrons). The biggest disadvantage that might complicate its physical realization is the requirement for a long coherence time of the quantum register. 

Another possibility is to initialize the second part of the quantum register at every iteration step (\textbf{B} version of IPEA). Every iteration step (not only the least important bits of $\phi$ as in Ref. \cite{dobsicek_2007}) must be repeated and measurement statistics performed. One could otherwise possibly mix bits belonging to different eigenvalues in different iterations and obtain an unphysical result. The biggest advantage of this approach is avoidance of the long coherence times and therefore potentially easier physical implementation. On the other hand, the biggest disadvantage is that no improving of the overlap between the actual state of the quantum register and the exact wave function occurs during the iterations and one must ``fight" the overlap problem at every iteration step. But as our simulations have shown and is discussed further, the situation concerning this overlap problem is in fact quite acceptable in practise. Small number of repetitions of each iteration is sufficient for amplification of the success probability to unity, when a suitable initial state of the quantum register is used.

\subsubsection*{Application to molecular Hamiltonians}
The relevance of phase estimation for quantum simulations was first noticed by Abrams and Lloyd \cite{abrams_1999}. They took $\hat{U}$ in the form

\begin{equation}
 \hat{U} = e^{i\tau\hat{H}},  
\end{equation}

\noindent
where $\hat{H}$ is the Born-Oppenheimer electronic Hamiltonian and $\tau$ is a suitable parameter which assures $\phi$ being in the interval $\langle 0,1)$.

The electronic Hamiltonian can be expressed in the second quantized form as \cite{szabo_ostlund}

\begin{equation}
 \hat{H} = \sum_{pq} h_{pq} \hat{a}_{p}^{\dagger} \hat{a}_{q} + \frac{1}{2} \sum_{pqrs} \langle pq | rs \rangle \hat{a}_{p}^{\dagger} \hat{a}_{q}^{\dagger} \hat{a}_{s} \hat{a}_{r} = \sum_{X=1}^{L} \hat{h}_{X},
  \label{ham_sec_quant}
\end{equation}

\noindent
where $h_{pq}$ and $\langle pq | rs \rangle$ are one and two-electron integrals in the molecular spin orbital basis and $\hat{a}_{i}^{\dagger}$ and $\hat{a}_{i}$ are fermionic creation and annihilation operators. The whole summation is formally expressed as a sum of individual terms $\hat{h}_{X}$. Since these creation and annihilation operators in general do not commute, exponential of the Hamiltonian cannot be written as a product of exponentials of individual $\hat{h}_{X}$, but a numerical approximation must be used \cite{lloyd_1996}. The first-order Trotter approximation \cite{trotter} can be expressed as

\begin{equation}
 e^{i\tau\hat{H}} = e^{i\tau\sum_{X=1}^{L}\hat{h}_{X}} = \Big(\prod_{X=1}^{L} e^{i\hat{h}_{X}\tau/N}\Big)^{N} + \mathcal{O}(\tau^{2}/N).
 \label{trotter}
\end{equation}

By choosing $N \geq (\tau^{2}/\epsilon)$, we can implement $\hat{U}$ within an error tolerance of $\mathcal{O}(\epsilon)$ using $\mathcal{O}(L(\tau^{2}/\epsilon))$ particular terms $e^{i\hat{h}_{X}\tau/N}$. Because $L$ scales as $\mathcal{O}(n^{4})$ in the total number of spin orbitals $n$ and individual terms $e^{i\hat{h}_{X}\tau/N}$ can be built efficiently \cite{ortiz_2001}, the whole algorithm is efficient (polynomial). 

When taking into account that about twenty binary digits of $\phi$ are necessary for quantum chemical applications, the maximal power of $\hat{U}$ in IPEA is $2^{(m-1)}=2^{19}\approx 5 \cdot 10^{5}$. This brings quite a big prefactor to the polynomial scaling of the algorithm and it is surely not necessary to emphasize that the Trotter approximation is the bottleneck of the algorithm.   

In our implementation, two external inputs are necessary. These are maximum ($E_{\mathrm{max}}$) and minimum ($E_{\mathrm{min}}$) energies expected in the studied system. We use $\hat{U}$ in the form

\begin{equation}
 \hat{U} = e^{i\tau(E_{\mathrm{max}} - \hat{H})},
 \label{our_u}
\end{equation}

\noindent
where $\tau$ reads

\begin{equation}
 \tau = \frac{2\pi}{E_{\mathrm{max}} - E_{\mathrm{min}}}
\end{equation}

\noindent
and the final energy is obtained according to the formula

\begin{equation}
 E = E_{\mathrm{max}}-\frac{2\pi\phi}{\tau}.
\end{equation}

This approach assures $\phi$ to be in the interval $\langle 0,1)$. Aspuru-Guzik variant \cite{aspuru-guzik_2005}, where $\phi = 0.5$ for Hartree-Fock energy ($E_{\mathrm{SCF}}$), corresponds to the choice of $E_{\mathrm{max}} = 0$ and $E_{\mathrm{min}} = 2E_{\mathrm{SCF}}$.

$E_{\mathrm{min}}$ and $E_{\mathrm{max}}$ can be chosen arbitrarily, but one must be sure that the calculated energy is within this interval, otherwise one would end up with a non-physical energy, due to the periodicity of $e^{2\pi i \phi}$. The smaller the interval between the minimum and the maximum energies is, the less iterations of IPEA are necessary for desired precision of $E$ and therefore less quantum gates are used. However, as the interval is smaller, $\tau$ is bigger and more repetitions of the Trotter approximation (\ref{trotter}) must be performed. In fact, when obtaining the energy with a fixed precision, the total exponential factor $\tau \cdot 2^{m-1}$ is for the most ``expensive" powers of $\hat{U}$ constant and independent of the size of the energy interval.

Taking $\hat{U}$ in the form (\ref{our_u}) does not pose any difficulties and indeed as the following circuit equality shows, just one more one-qubit rotation is needed.

\begin{center}
\mbox{
 \Qcircuit @C=1em @R=0.5em {
   & \ctrl{2} & \qw & & & \ctrl{2} & \gate{\left(\begin{array}{cc} 
                                         1 & 0 \\
                                         0 & e^{i\tau E_{\mathrm{max}}}
                                         \end{array}\right)} & \qw \\
   & & & = & & \\                                      
   & \gate{e^{i\tau (E_{\mathrm{max}} - \hat{H})}} & \qw & & & \gate{e^{-i\tau \hat{H}}} & \qw & \qw
 }
}
\end{center}

\subsection{Factorization of the exponential of a second quantized Hamiltonian}
Since we did not perform the factorization of the exponential of the molecular electronic Hamiltonian to one and two-qubit elementary gates in the numerical simulations, we will only briefly mention this topic, that was first studied by Ortiz et al. \cite{ortiz_2001}. 

The first step is the Jordan-Wigner transformation \cite{jordan_1928} which maps the fermionic creation and annihilation operators to spin operators represented by Pauli $\sigma$-matrices. The transformation has the form

\begin{equation}
 \hat{a}_{n}^{\dagger} = \Bigg( \prod_{j=1}^{n-1} \sigma_{z}^{j} \Bigg) \sigma_{+}^{n}, \quad \hat{a}_{n} = \Bigg( \prod_{j=1}^{n-1} \sigma_{z}^{j} \Bigg) \sigma_{-}^{n},
\end{equation}

\noindent
where $\sigma_{\pm} = 1/2(\sigma_{x} \pm i \sigma_{y})$ and the superscript denotes the qubit on which the matrix acts. After application of this transformation,  the molecular Hamiltonian can be written in terms of strings of $\sigma$-matrices. It was shown that individual exponentials containing these strings can be built efficiently from one and two-qubit elementary gates \cite{ortiz_2001, ovrum_2007, lanyon_2010, whitfield_2010}. The circuit representations of individual terms can be found in \cite{whitfield_2010} or in the Supplementary Information of \cite{lanyon_2010}.

Once the factorization scheme is known, one can study the complexity of the whole algorithm. Scaling of the algorithm is given by the scaling of a single controlled action of the unitary operator $\Big(\prod_{X=1}^{L} e^{i\hat{h}_{X}\tau/N}\Big)$ from equation (\ref{trotter}). Repetitions in the Trotter approximation (\ref{trotter}) increase only the prefactor to the polynomial scaling, not the scaling itself. Also the required precision is limited, about twenty binary digits of $\phi$ are sufficient for chemical accuracy. Detailed complexity analysis presented in \cite{ovrum_2007} revealed that the algorithm scales as $\mathcal{O}(n_{\mathrm{sorb}}^{5})$ in the number of spin orbitals (number of qubits in the second part of the quantum register). This is in contrast to the exponential scaling of the full configuration interaction method on a conventional computer. 

\begin{figure}[!h]
 \begin{center}
   \includegraphics[width=12cm]{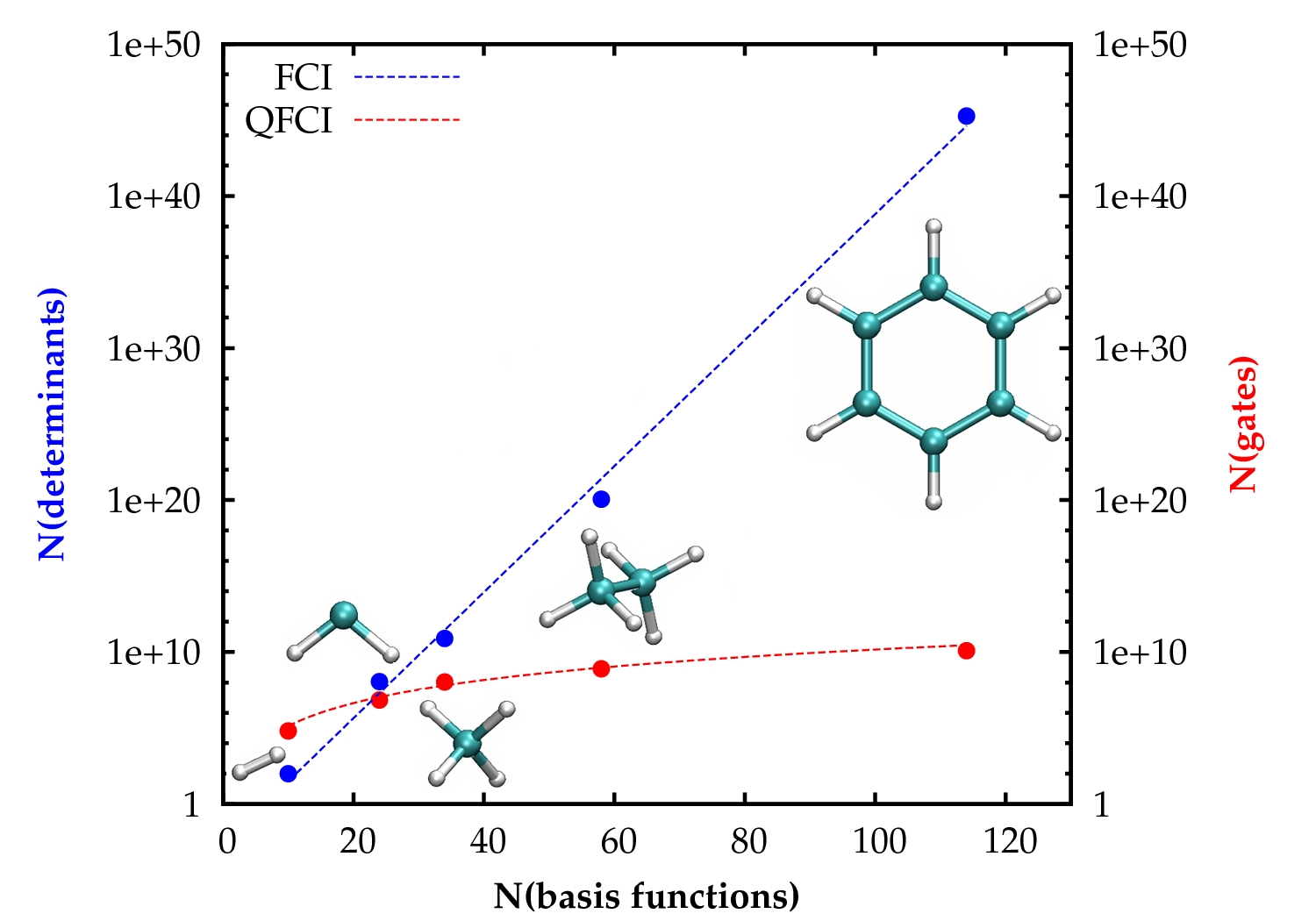}
   \caption{The exponential speedup of the QFCI over the FCI. In case of the FCI (blue), dependence of the number of Slater determinants in the FCI expansion on the number of basis functions is shown. In case of the QFCI (red), dependence of the number of one and two-qubit gates needed for a single controlled action of the unitary operator (for details see the text) on the number of basis functions is presented. There are logarithmic scales on $y$ axes and the points in the graph correspond to depicted molecules (hydrogen, methylene, methane, ethane, and benzene) in the cc-pVDZ basis set.} 
   \label{scaling}    
 \end{center} 
\end{figure}

The exponential speedup is demonstrated in Figure \ref{scaling}. Quantum gates count for a single controlled action of the unitary operator was done according to \cite{whitfield_2010} and the number of gates corresponds to standard one and two-qubit gates from the set: Hadamard gate (one-qubit), CNOT gate (two-qubit), single qubit $x$-rotation $R_{x}(-\pi/2)$, single qubit $z$-rotation $R_{z}(\theta)$ and controlled $R_{z}(\theta)$ (two-qubit). We have to emphasize at this point that a simulation of the QFCI method on a conventional computer \textit{cannot} achieve this exponential speedup in principle. Even though only polynomial number of elementary quantum gates is necessary, the simulation of an action of such single elementary gate scales exponentially itself, since it corresponds to a matrix vector multiplication of the dimension $2^{n}$.

Here we would like to note three more things. Firstly, we assumed that initial state preparation is an efficient step, as was already mentioned. Secondly, when a quantum chemical method with a scaling worse than $\mathcal{O}(n^5)$ is used for calculation of an initial guess state on a conventional computer, then this classical step is becoming a rate determining one. Besides this, the classical computation of the integrals in the molecular orbital basis scales as $\mathcal{O}(n^{5})$ (due to the integral transformation) as well.




\section{Application to four lowest-lying electronic states of CH$_{2}$}
Methylene molecule (CH$_{2}$) in a minimal basis set (STO-3G) is a simple, yet computationally interesting system suitable for simulations and testing of the aforementioned QFCI algorithm. CH$_{2}$ molecule is well known for the multireference character of its lowest-lying singlet electronic state ($\tilde{a}~^{1}A_{1}$) and is often used as a benchmark system for testing of newly developed computational methods (see e.g. \cite{bwcc1, evangelista-allen1, mkcc_our2, demel-pittner-bwccsdt}). When using the STO-3G basis set, the total number of molecular (spin)orbitals is 7(14). We therefore work with 15 qubits in the direct mapping approach (one qubit is needed in the read-out part of the register). In spite of the fact that the complexity of simulations of the QFCI on conventional computers scales exponentially as the complexity of the classical FCI but with an order of magnitude larger prefactor \cite{aspuru-guzik_2005}, this system is still computationally feasible and for its properties an excellent candidate for one of the first benchmark simulations.

Our aim was to verify the applicability of the QFCI for the ground as well as excited states exhibiting multireference character. We accordingly simulated the QFCI energy calculations of the four lowest-lying electronic states of CH$_{2}$: $\tilde{X}~^{3}B_{1}$, $\tilde{a}~^{1}A_{1}$, $\tilde{b}~^{1}B_{1}$, and $\tilde{c}~^{1}A_{1}$. For CH$_{2}$ at the equilibrium geometry, the ground electronic state is not a closed-shell singlet, but a triplet state ($\tilde{X}~^{3}B_{1}$) with the electronic configuration

\begin{equation}
 (1a_{1})^{2}(2a_{1})^{2}(1b_{2})^{2}(3a_{1})(1b_{1}).
 \label{conf1}
\end{equation}

\noindent
The closed-shell singlet state ($\tilde{a}~^{1}A_{1}$), which can be qualitatively described by the electronic configuration

\begin{equation}
 (1a_{1})^{2}(2a_{1})^{2}(1b_{2})^{2}(3a_{1})^{2},
 \label{conf2}
\end{equation}

\noindent
is the first excited state. This state exhibits the multireference character with the second important configuration

\begin{equation}
 (1a_{1})^{2}(2a_{1})^{2}(1b_{2})^{2}(1b_{1})^{2}.
 \label{conf3}
\end{equation}

\noindent
The contribution of both closed-shell configurations becomes equal at linear geometries. The third electronic state ($\tilde{b}~^{1}B_{1}$) has the same spatial orbital configuration (\ref{conf1}) as the ground state, but with singlet-coupled open shell electrons. The fourth electronic state ($\tilde{c}~^{1}A_{1}$) is represented by the same two configurations as the $\tilde{a}~^{1}A_{1}$ state but the amplitudes have the same sign and the amplitude of (\ref{conf3}) is greater than that of (\ref{conf1}) [this state can be qualitatively described by the configuration (\ref{conf3})].

\begin{figure}[!h]
 \begin{center}
   \includegraphics[width=12cm]{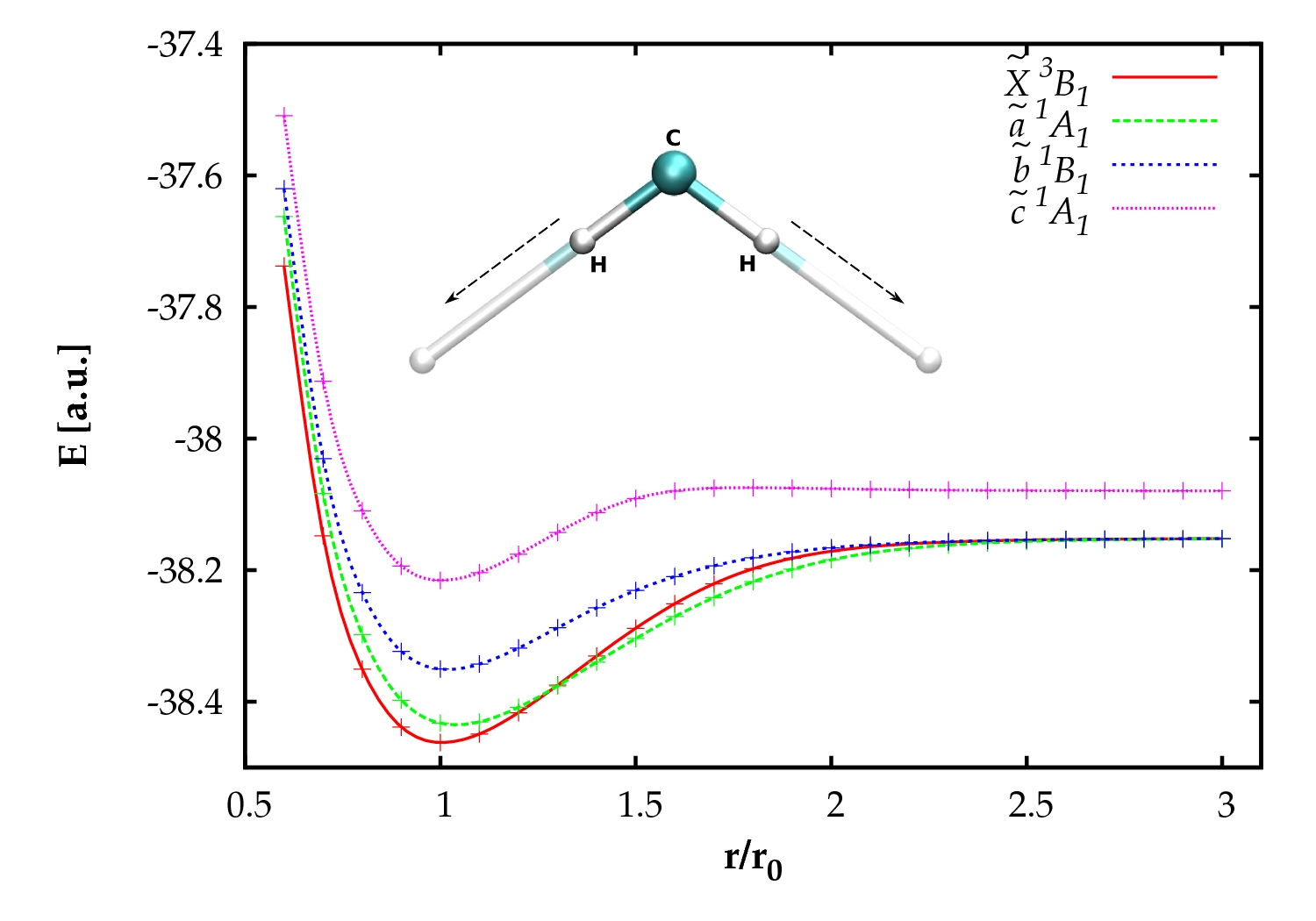}
   \caption{Energies of the four simulated states of CH$_{2}$ for the C-H bond stretching, $r_{0}$ denotes the equilibrium bond distance.} 
   \label{stretch}    
 \end{center} 
\end{figure}

\begin{figure}[!h]
 \begin{center}
   \includegraphics[width=12cm]{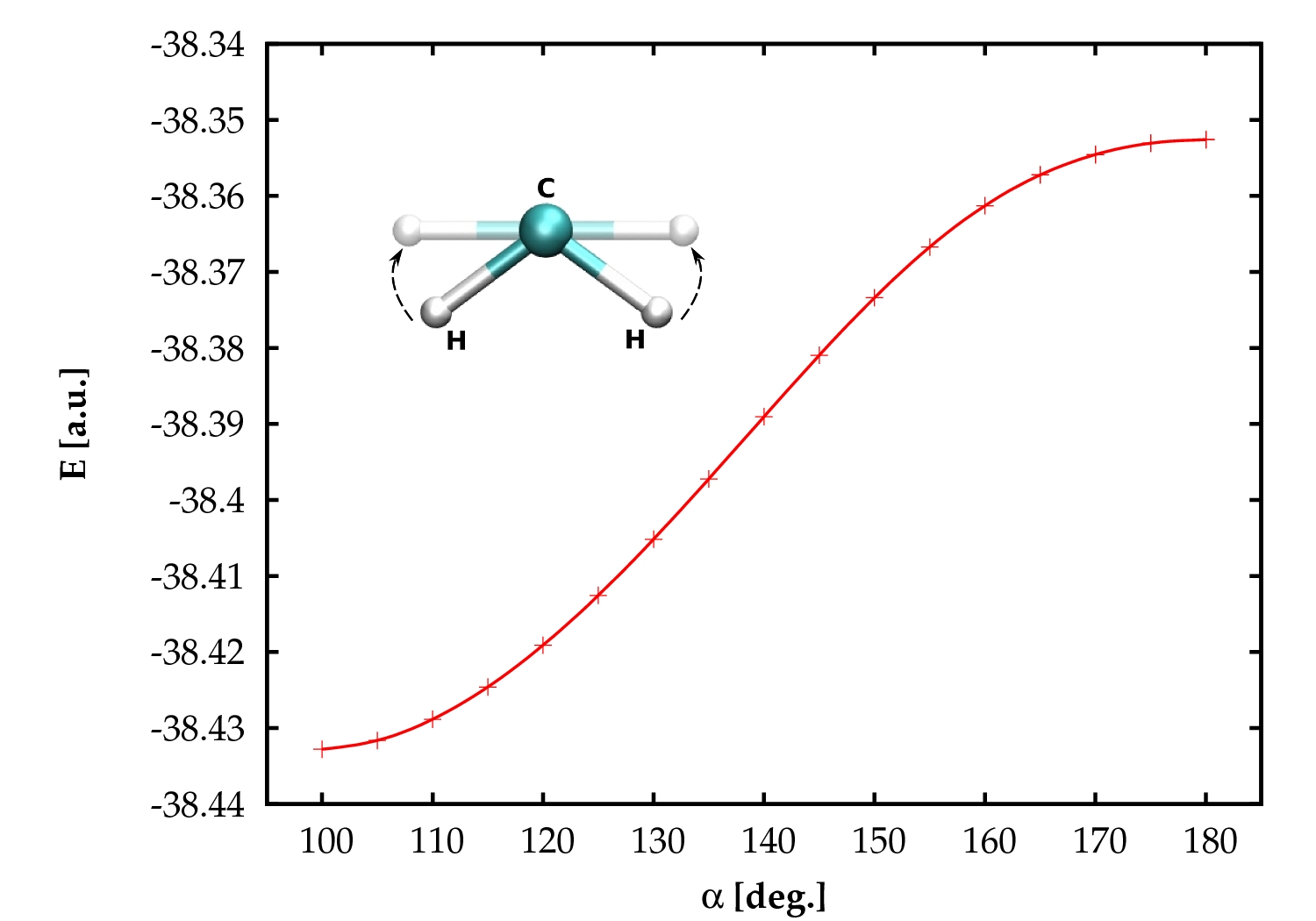}
   \caption{Energy of $\tilde{a}~^{1}A_{1}$ state of CH$_{2}$ for the H-C-H angle bending, $\alpha$ denotes the H-C-H angle.} 
   \label{bend}    
 \end{center} 
\end{figure}

We simulated the QFCI energy calculations for C-H bond stretching (both C-H bonds were stretched, Figure \ref{stretch}), and H-C-H angle bending for $\tilde{a}~^{1}A_{1}$ state (Figure \ref{bend}). These processes were chosen designedly because description of bond breaking is a hard task for many of computational methods and H-C-H angle bending since the $\tilde{a}~^{1}A_{1}$ state exhibits very strong multireference character at linear geometries. The equilibrium geometry of CH$_{2}$ molecule was adopted from \cite{sherrill_1997} and corresponded to $r_{e} = 1.1089~\rm{\AA}$ and $\alpha_{e} = 101.89^{\circ}$). Our work follows up the work by Wang et al. \cite{wang_2008}. In this paper, the authors studied the influence of initial guesses on the performance of the quantum FCI method on two singlet states of water molecule across the bond-dissociation regime. They found out that the Hartree-Fock initial guess is not sufficient for bond dissociation and suggested the use of MCSCF method (CASSCF in particular). Few configuration state functions added to the initial guess improved the success probability dramatically.

We also used and tested different initial guesses for QFCI calculations. Those denoted as HF guess were composed only from spin-adapted configurations which qualitatively describe certain state: in case of $\tilde{a}~^{1}A_{1}$ configuration (\ref{conf2}), in case of $\tilde{c}~^{1}A_{1}$ configuration (\ref{conf3}), in case of $\tilde{X}~^{3}B_{1}$ two triplet-coupled configurations (\ref{conf1}) (with weights 1/2) and for $\tilde{b}~^{1}B_{1}$ the same two configurations but singlet-coupled. Initial guesses denoted as CAS($x$,$y$) guess were based on CASCI calculations with small complete active spaces (more details about the definition of the active spaces will be given further), which contained $x$ electrons in $y$ orbitals. Initial guesses based on CASSCF calculations as in \cite{wang_2008} could be used in the same way. To be consistent, we employed the FCI wave functions in a limited active space composed of RHF orbitals, which were also used for the exponential of a Hamiltonian in the QFCI algorithm. Initial guesses were constructed only from the configurations whose absolute values of amplitudes were higher than 0.1. Those constructed from the configurations whose absolute values of amplitudes were higher than 0.2 are denoted as CAS($x$,$y$), tresh. 0.2 guess. All the initial guesses were normalized before the simulations.

Similarly as in Ref. \cite{aspuru-guzik_2005}, the exponential of a Hamiltonian operator was implemented as a $n$-qubit gate. Factorization to the elementary one and two-qubit gates was performed only to examine the gate count, but not in the numerical simulations. We also did not take into account any decoherence and thus assumed that the exponential of the Hamiltonian can be obtained with an arbitrary precision by a proper number of repetitions in (\ref{trotter}). One and two-electron integrals in the MO basis, parametrizing the Hamiltonian (\ref{ham_sec_quant}), were obtained using the restricted Hartree-Fock (RHF) orbitals. All \textit{ab initio} calculations (FCI, RHF) were employed with our suite of quantum chemical programs \cite{gelfand}.

The phase in IPEA was always computed up to $m=20$ binary digits. Maximum and minimum expected energies needed for the algorithm were set to $E_{\mathrm{max}} = -37.5$ a.u. and $E_{\mathrm{min}} = -39.0$ a.u. All presented success probabilities correspond to sum of the probabilities of rounding the phase up and down ($P_{\mathrm{tot}} = P_{\mathrm{up}} + P_{\mathrm{down}}$), therefore to  probabilities of obtaining the final energy with precision $\approx 1.43 \cdot 10^{-6}$ a.u.  

Finally, both of the aforementioned variants of IPEA (\textbf{A} and \textbf{B}) were tested.

\section{Results}
\subsection{C-H bond stretching}
\begin{figure}[!h]
\begin{minipage}{0.45\linewidth}
 \begin{center}
   \includegraphics[width=7.5cm]{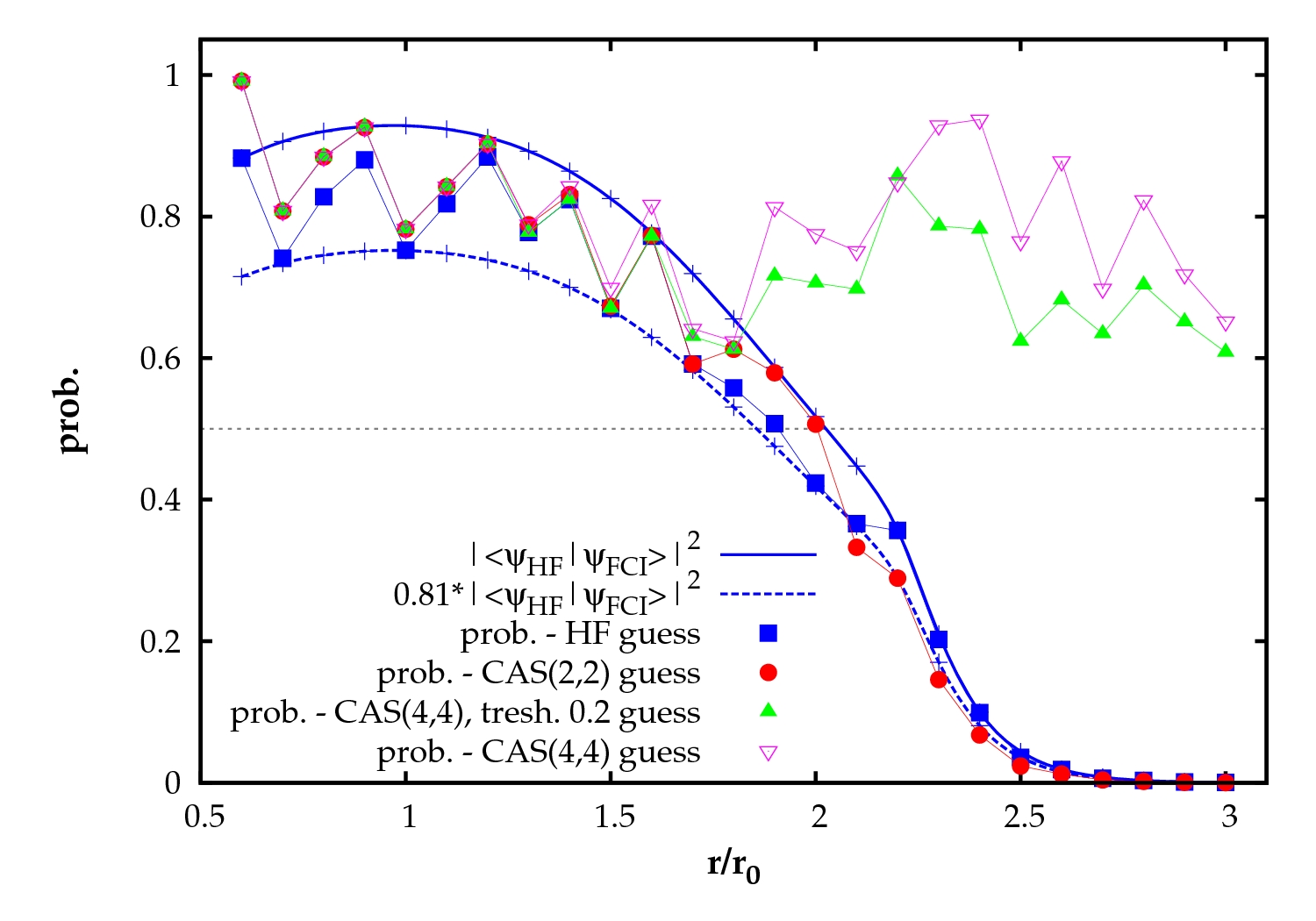}    
   a) $\tilde{a}~^{1}A_{1}$ state  
 \end{center} 
\end{minipage}  
\begin{minipage}{0.45\linewidth}
 \begin{center}
   \includegraphics[width=7.5cm]{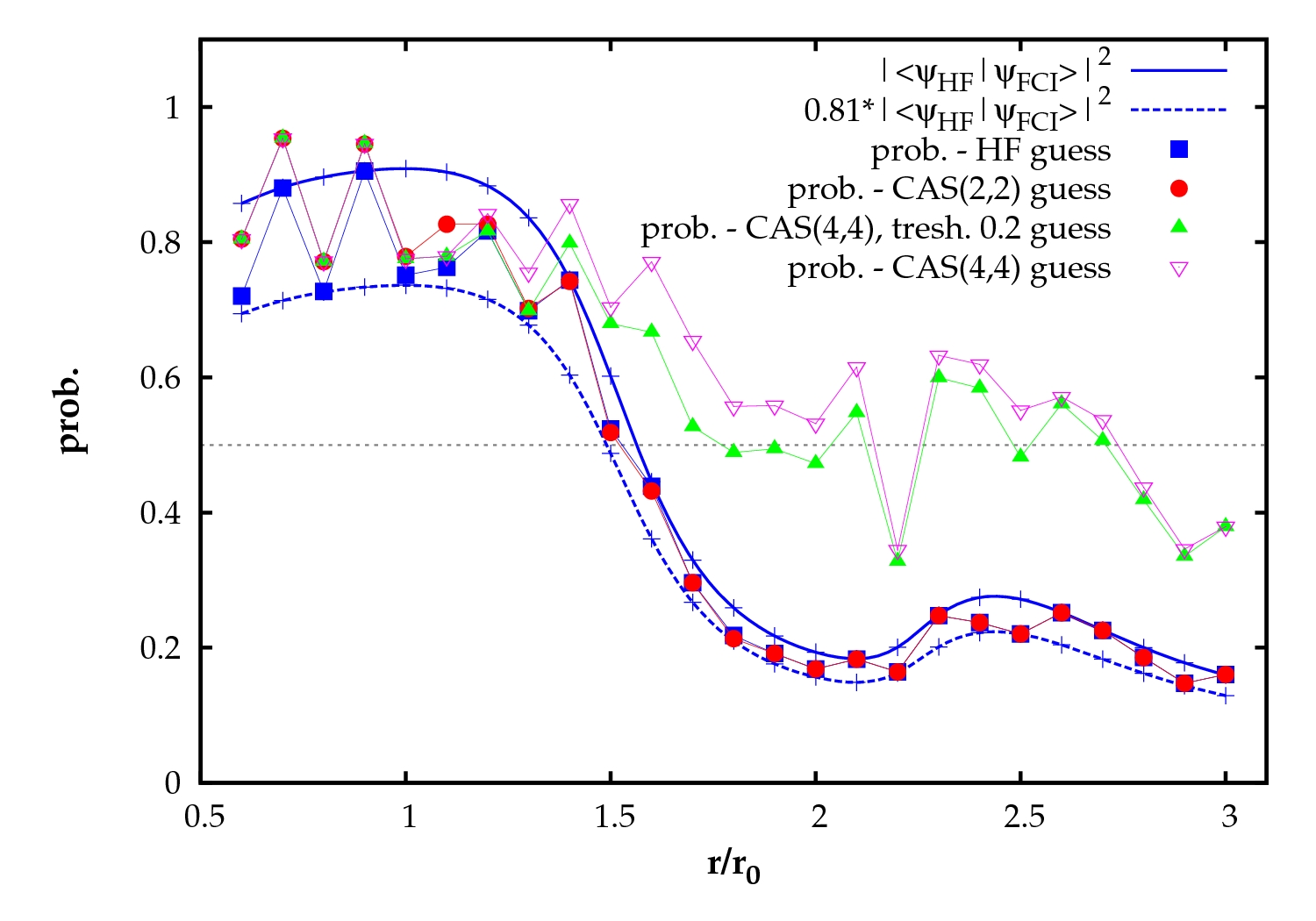}    
   b) $\tilde{c}~^{1}A_{1}$ state  
 \end{center} 
\end{minipage}  

\vskip 0.3cm  

\begin{minipage}{0.45\linewidth}
 \begin{center}
   \includegraphics[width=7.5cm]{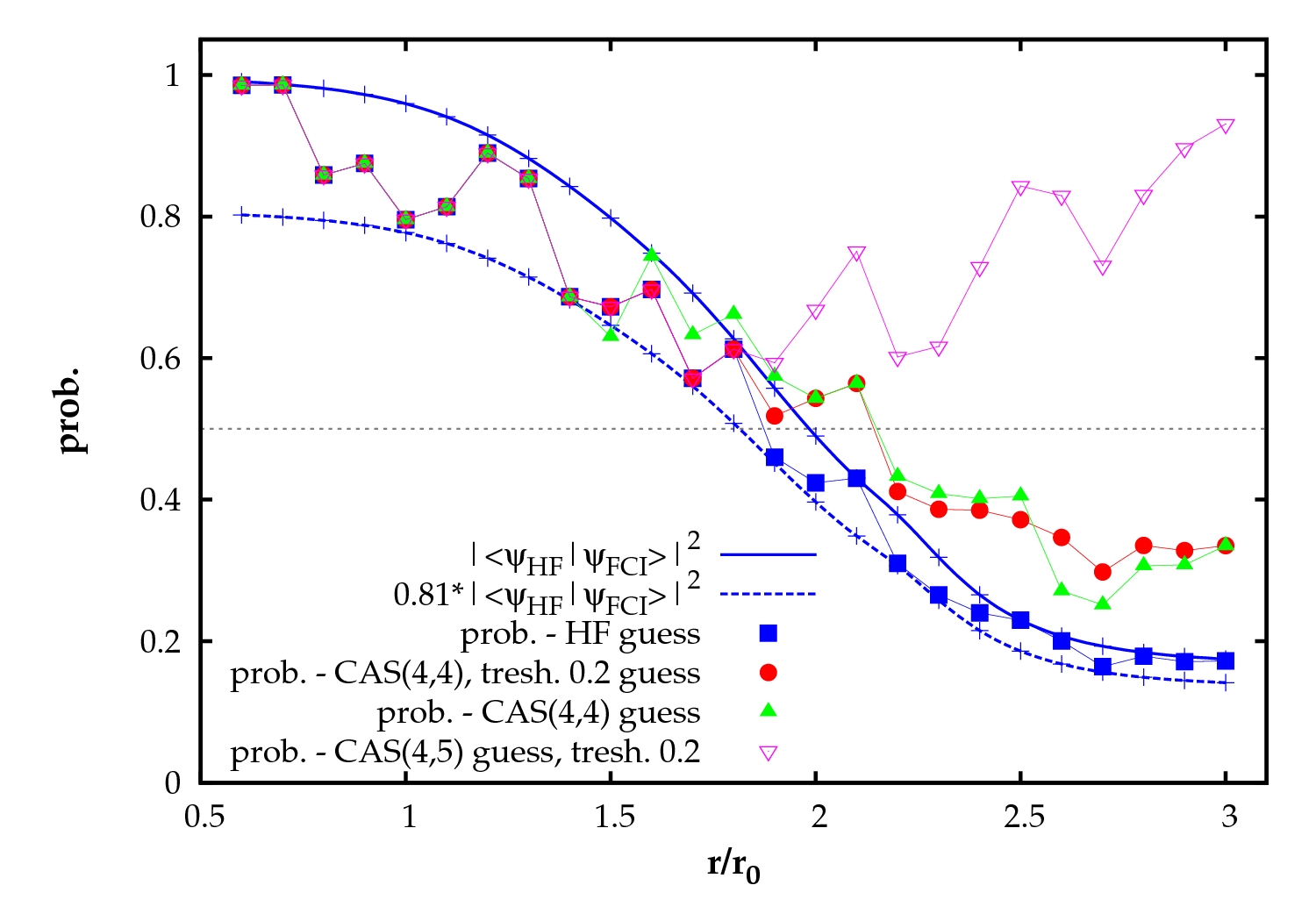}    
   c) $\tilde{X}~^{3}B_{1}$ state  
 \end{center} 
\end{minipage}  
\begin{minipage}{0.45\linewidth}
 \begin{center}
   \includegraphics[width=7.5cm]{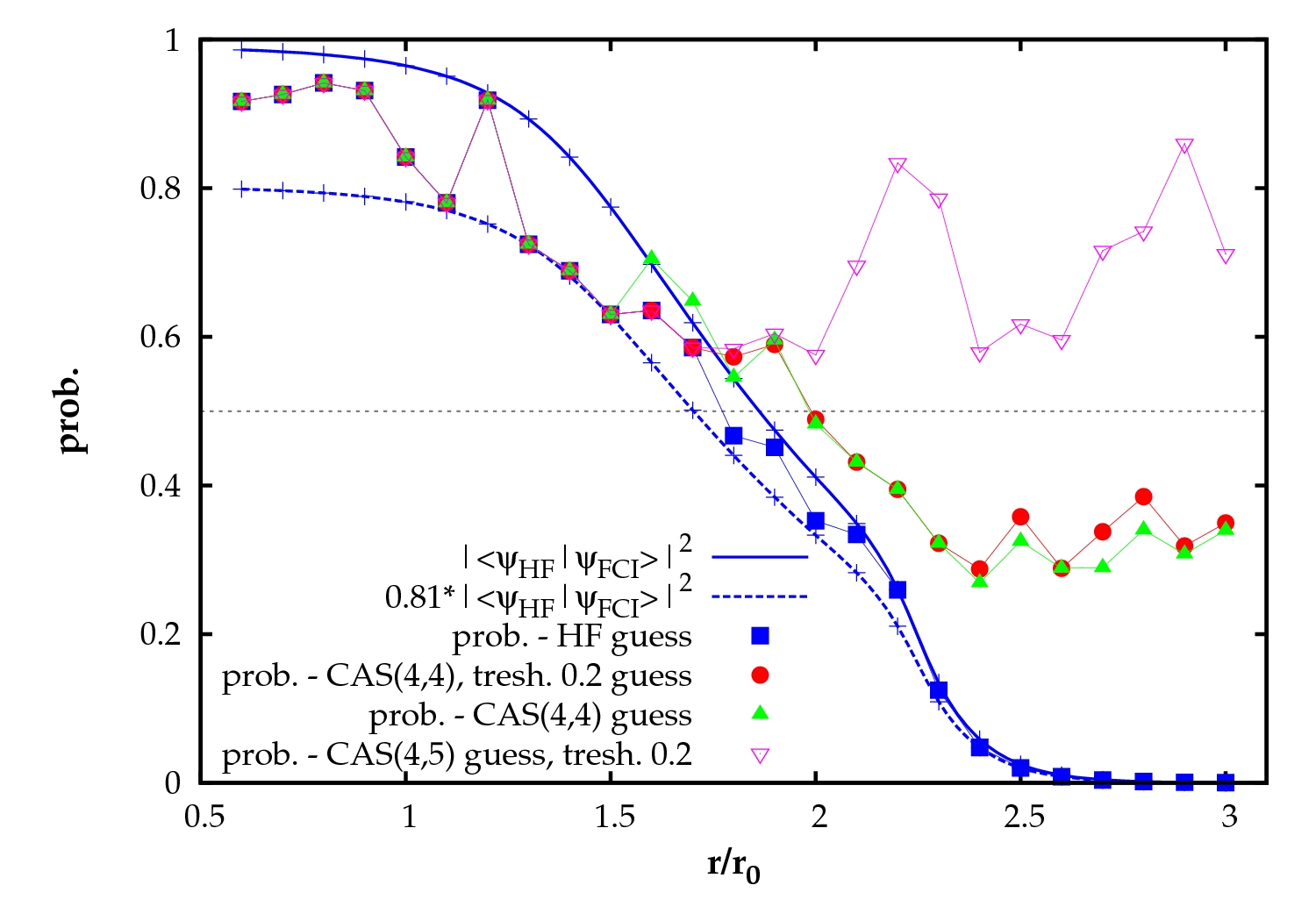}    
   d) $\tilde{b}~^{1}B_{1}$ state  
 \end{center} 
\end{minipage} 
\caption{Success probabilities of the \textbf{A} version of IPEA for the four electronic states of CH$_{2}$ and different initial guesses, tresh 0.2 means that only configurations with absolute values of amplitudes higher than 0.2 were involved in the initial guess, $r_{0}$ denotes the equilibrium bond distance.}
\label{stretch_alg0}
\end{figure}

\begin{figure}[!h]
 \begin{center}
   \includegraphics[width=12cm]{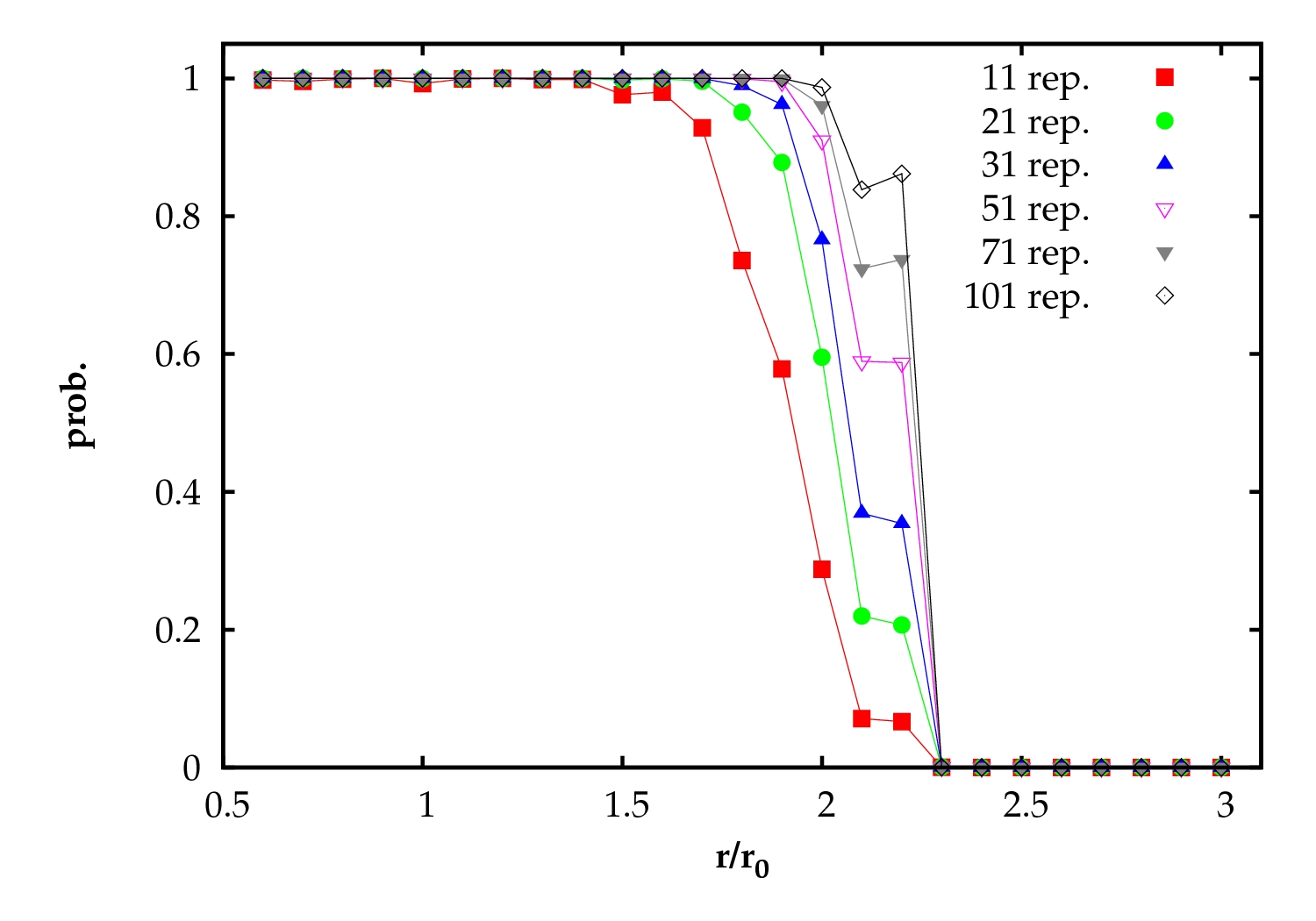}
   \caption{Success probabilities of the \textbf{B} version of IPEA with HF guess for $\tilde{a}~^{1}A_{1}$ state and different number of repetitions of individual bit measurements, $r_{0}$ denotes the equilibrium bond distance.} 
   \label{singlet_a1_lower_rep_hf}    
 \end{center} 
\end{figure}

\begin{figure}[!h]
\begin{minipage}{0.45\linewidth}
 \begin{center}
   \includegraphics[width=7.5cm]{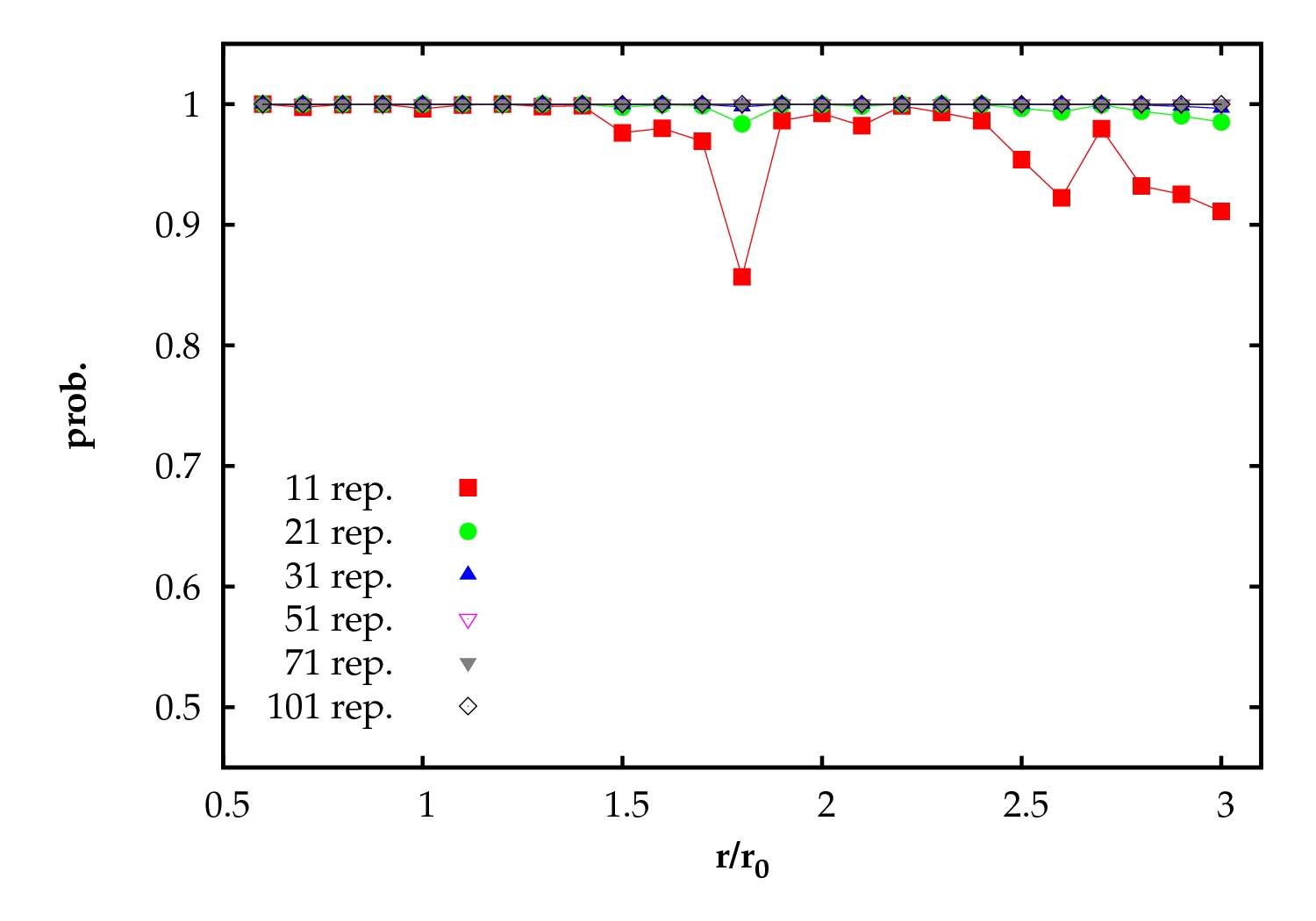}    
   a) $\tilde{a}~^{1}A_{1}$ state, CAS(4,4), tresh 0.2 guess
 \end{center} 
\end{minipage}
\begin{minipage}{0.45\linewidth}
 \begin{center}
   \includegraphics[width=7.5cm]{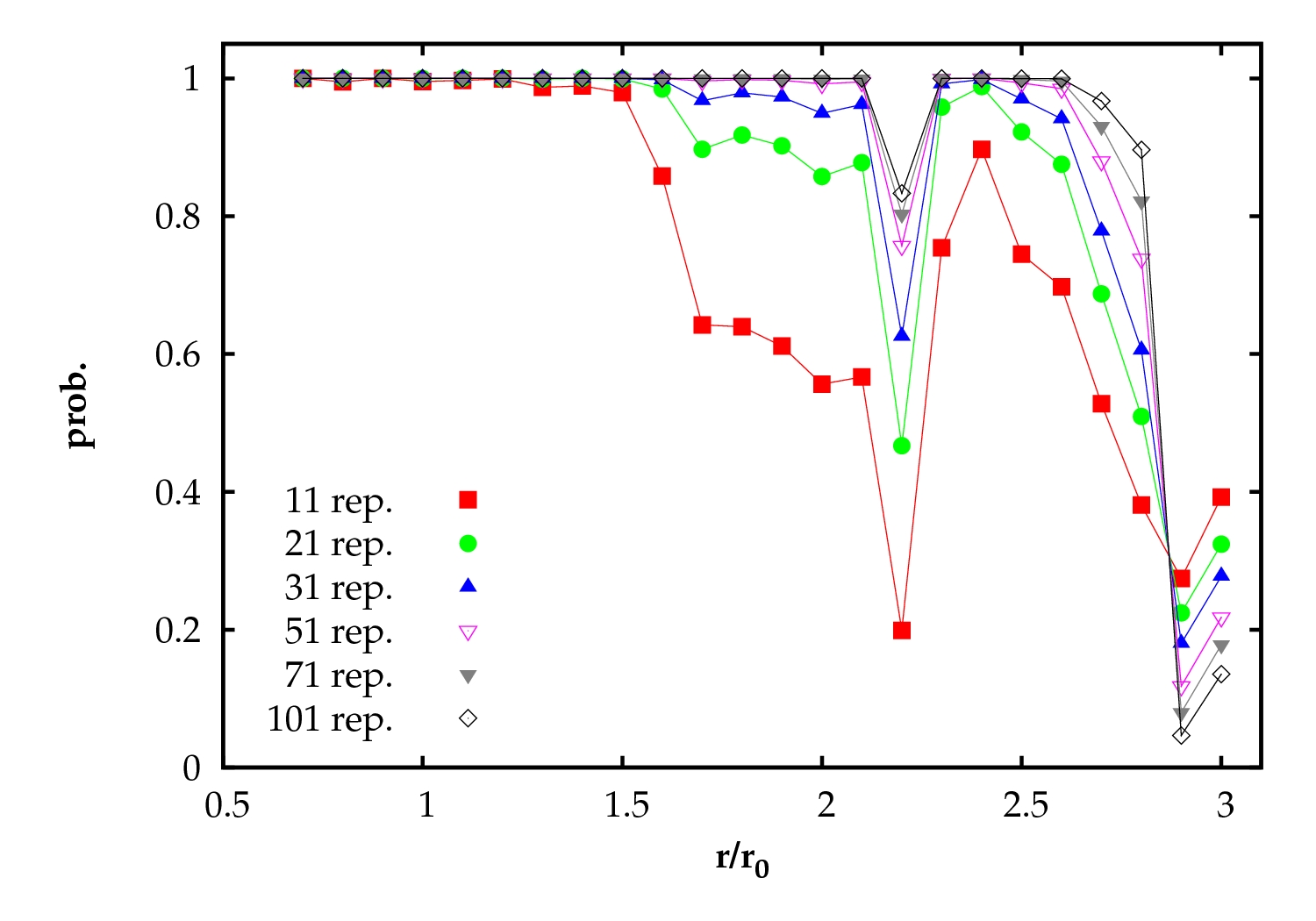}    
   b) $\tilde{c}~^{1}A_{1}$ state, CAS(4,4), tresh 0.2 guess
 \end{center} 
\end{minipage}

\vskip 0.3cm  

\begin{minipage}{0.45\linewidth}
 \begin{center}
   \includegraphics[width=7.5cm]{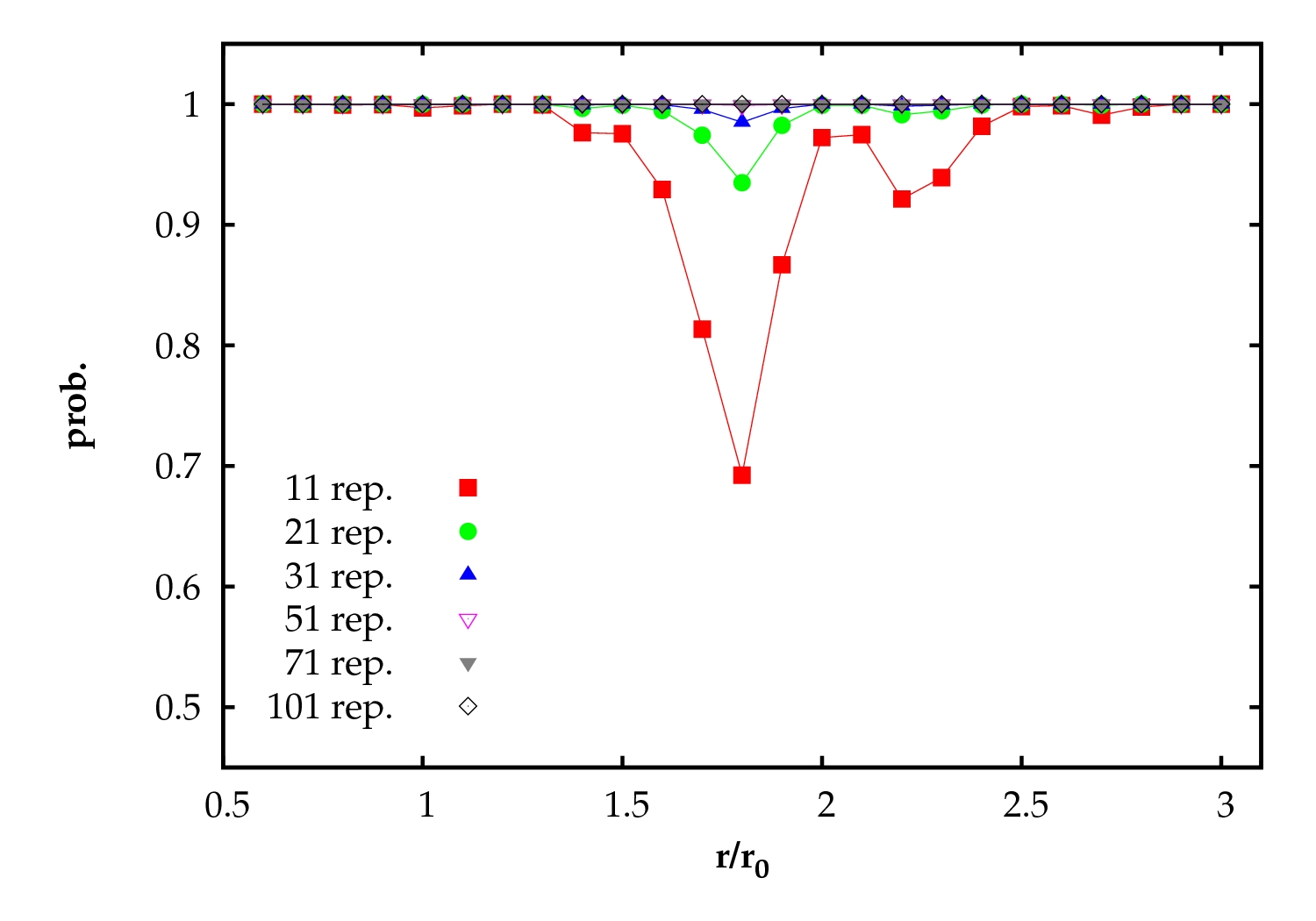}    
   c) $\tilde{X}~^{3}B_{1}$ state, CAS(4,5), tresh 0.2 guess
 \end{center} 
\end{minipage}
\begin{minipage}{0.45\linewidth}
 \begin{center}
   \includegraphics[width=7.5cm]{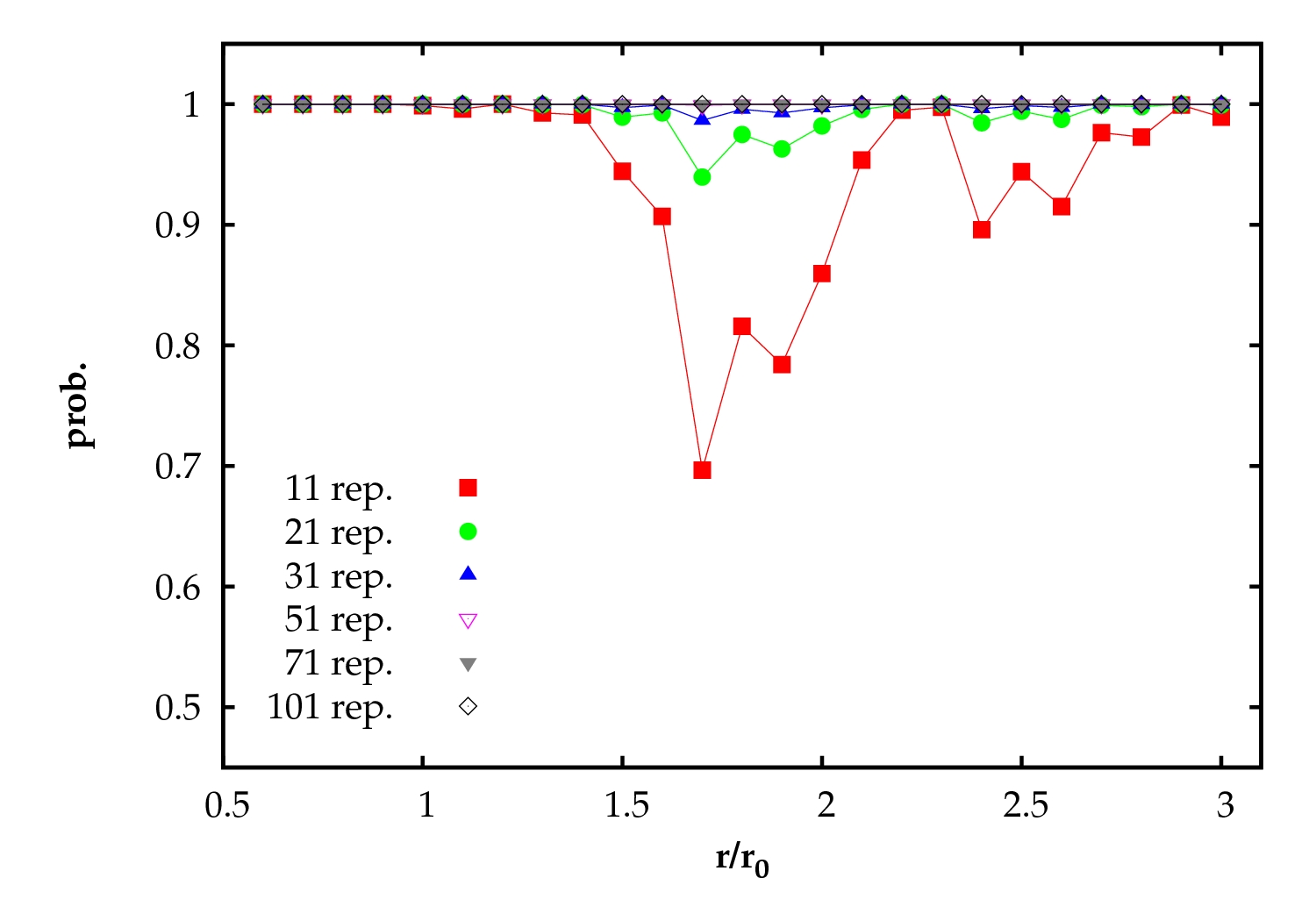}    
   d) $\tilde{b}~^{1}B_{1}$ state, CAS(4,5), tresh 0.2 guess
 \end{center} 
\end{minipage}
\caption{Success probabilities of the \textbf{B} version of IPEA with ``best" initial guesses and different number of repetitions of individual bit measurements for all four states, $r_{0}$ denotes the equilibrium bond distance.}   
\label{stretch_alg1}
\end{figure}

Results for the C-H bond stretching are summarized in Figures \ref{stretch_alg0} - \ref{stretch_alg1}. 

Figure \ref{stretch_alg0} presents the performance of the \textbf{A} version of IPEA with maintaining the second part of the quantum register during all iterations. Subfigures a - d represent the simulations of the energy calculations of the four electronic states: a: $\tilde{a}~^{1}A_{1}$, b: $\tilde{c}~^{1}A_{1}$, c: $\tilde{X}~^{3}B_{1}$, and d: $\tilde{b}~^{1}B_{1}$. Overlap between the initial HF guess wave function and the exact FCI wave function as well as this overlap scaled by the factor 0.81 [according to (\ref{sc_factor})] are shown. Figure \ref{stretch_alg0} also presents the success probabilities of IPEA for the HF initial guess and initial guesses based on the CASCI calculations with certain small complete active spaces. Definition of these active spaces is complicated by the fact that swapping of molecular orbitals occurs when the C-H bonds are prolonged. To maximize the overlap between the initial and the exact wave functions, we constructed the active spaces
 from the actual highest occupied and the lowest unoccupied molecular orbitals at a given geometry. For $\tilde{X}~^{3}B_{1}$, $\tilde{b}~^{1}B_{1}$ and $\tilde{c}~^{1}A_{1}$, where the $1b_{1}$ orbital is involved in the qualitative description of the state (at the equilibrium geometry), this orbital was always included in the active space [$1b_{1}$ orbital which is the LUMO (5th molecular orbital) at the equilibrium geometry becomes the 7th when going to three times prolonged C-H bonds]. Definition of the complete active spaces is summarized in Table \ref{cas}. Dotted line corresponding to the probability 0.5 bounds the region where the algorithm can be safely used and the total probability amplified by repeating the whole process.

\begin{table}[t]
\begin{center}
\begin{small}
 \begin{tabular}{| c | l | l | l |}
 \hline
 \textbf{state} & \textbf{CAS(2,2)} & \textbf{CAS(4,4)} & \textbf{CAS(4,5)} \\
 \hline
 \multirow{2}{*}{$\tilde{a}~^{1}A_{1}$} & highest occupied MO, & two highest occupied MOs, & \\
  & lowest unoccupied MO & two lowest unoccupied MOs & \\
 \hline  
 \multirow{4}{*}{$\tilde{c}~^{1}A_{1}$} & highest occupied MO, & two highest occupied MOs, & \\
  & unoccupied $1b_{1}$ MO & unoccupied $1b_{1}$ MO, & \\
  & & lowest unoccupied MO & \\
  & & (other than $1b_{1}$) & \\  
 \hline  
 \multirow{4}{*}{$\tilde{X}~^{3}B_{1}$} & & two highest occupied MOs, & two highest occupied MOs, \\
  & & unoccupied $1b_{1}$ MO, & three lowest unoccupied MOs \\
  & & lowest unoccupied MO & (including $1b_{1}$) \\
  & & (other than $1b_{1}$) & \\
 \hline  
 \multirow{4}{*}{$\tilde{b}~^{1}B_{1}$} & & two highest occupied MOs, & two highest occupied MOs, \\
  & & unoccupied $1b_{1}$ MO, & three lowest unoccupied MOs \\
  & & lowest unoccupied MO & (including $1b_{1}$) \\
  & & (other than $1b_{1}$) & \\
 \hline  
 \end{tabular}
\caption{Summary of the complete active spaces (CAS) used for the calculations of initial guesses for IPEA (Figures \ref{stretch_alg0} and \ref{stretch_alg1}), occupation/unoccupation refers to the lowest closed-shell configuration (\ref{conf2}).}
\label{cas} 
\end{small} 
\end{center}
\end{table}

Figures \ref{singlet_a1_lower_rep_hf} and \ref{stretch_alg1} present the performance of the \textbf{B} version of IPEA. In this version, the second part of the quantum register is reinitialized at every iteration step. Figures \ref{singlet_a1_lower_rep_hf} and \ref{stretch_alg1} demonstrate the success probabilities for different number of such repetitions (11-101). Figure \ref{singlet_a1_lower_rep_hf} shows the results and limits of the HF guess for $\tilde{a}~^{1}A_{1}$ state. Figure \ref{stretch_alg1} presents the results of the ``best" initial guesses in terms of price/performance ratio for all four states.   

\subsection{H-C-H angle bending}
\begin{figure}[!h]
 \begin{center}
   \includegraphics[width=12cm]{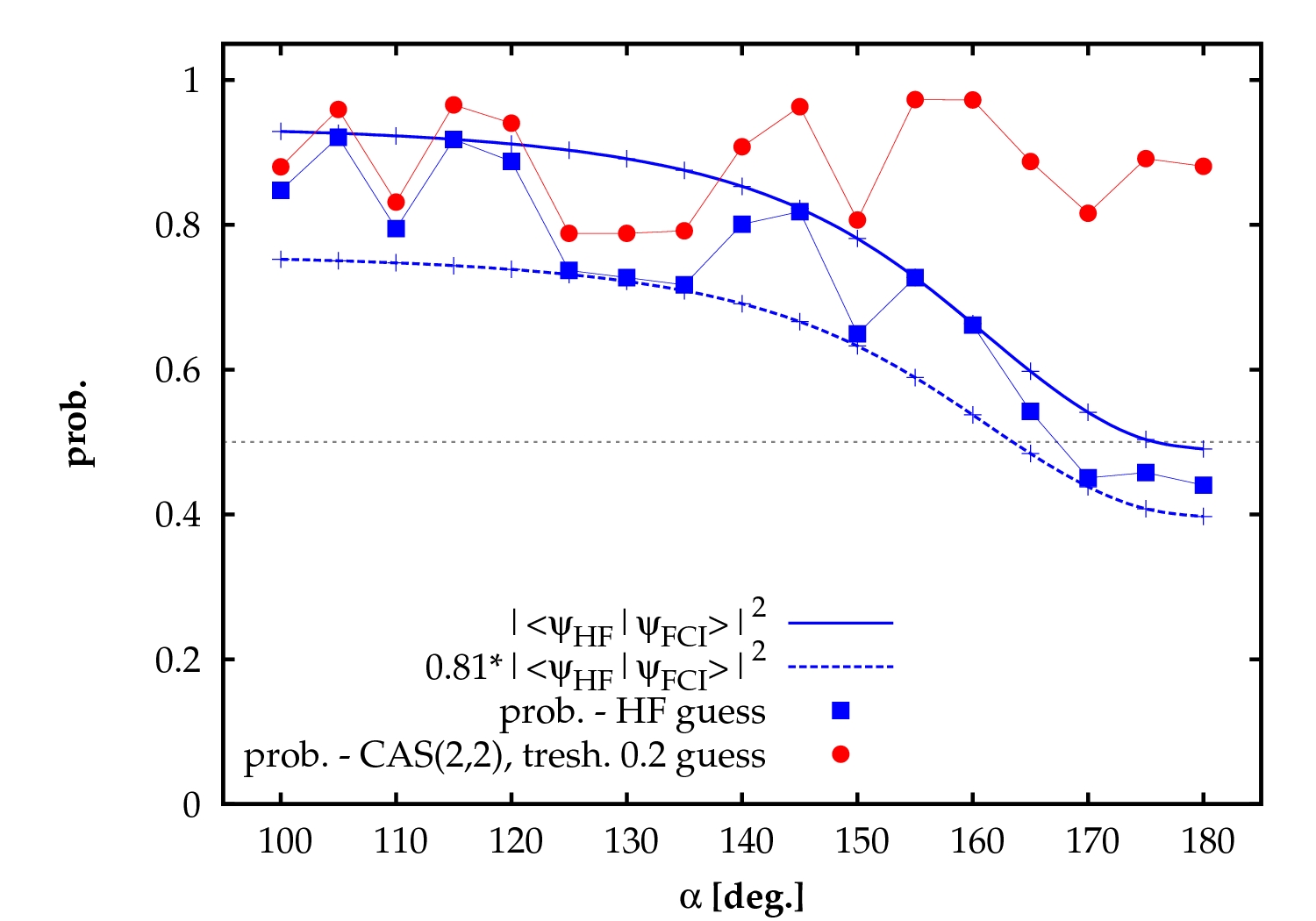}
   \caption{Success probabilities of the \textbf{A} version of IPEA for the $\tilde{a}~^{1}A_{1}$ state with HF and CAS(2,2) initial guesses, tresh 0.2 means that only configurations with absolute values of amplitudes higher than 0.2 were involved in the initial guess, $\alpha$ denotes the H-C-H angle.} 
   \label{bend_alg0}    
 \end{center} 
\end{figure}

\begin{figure}[!h]
\begin{minipage}{0.45\linewidth}
 \begin{center}
   \includegraphics[width=7.5cm]{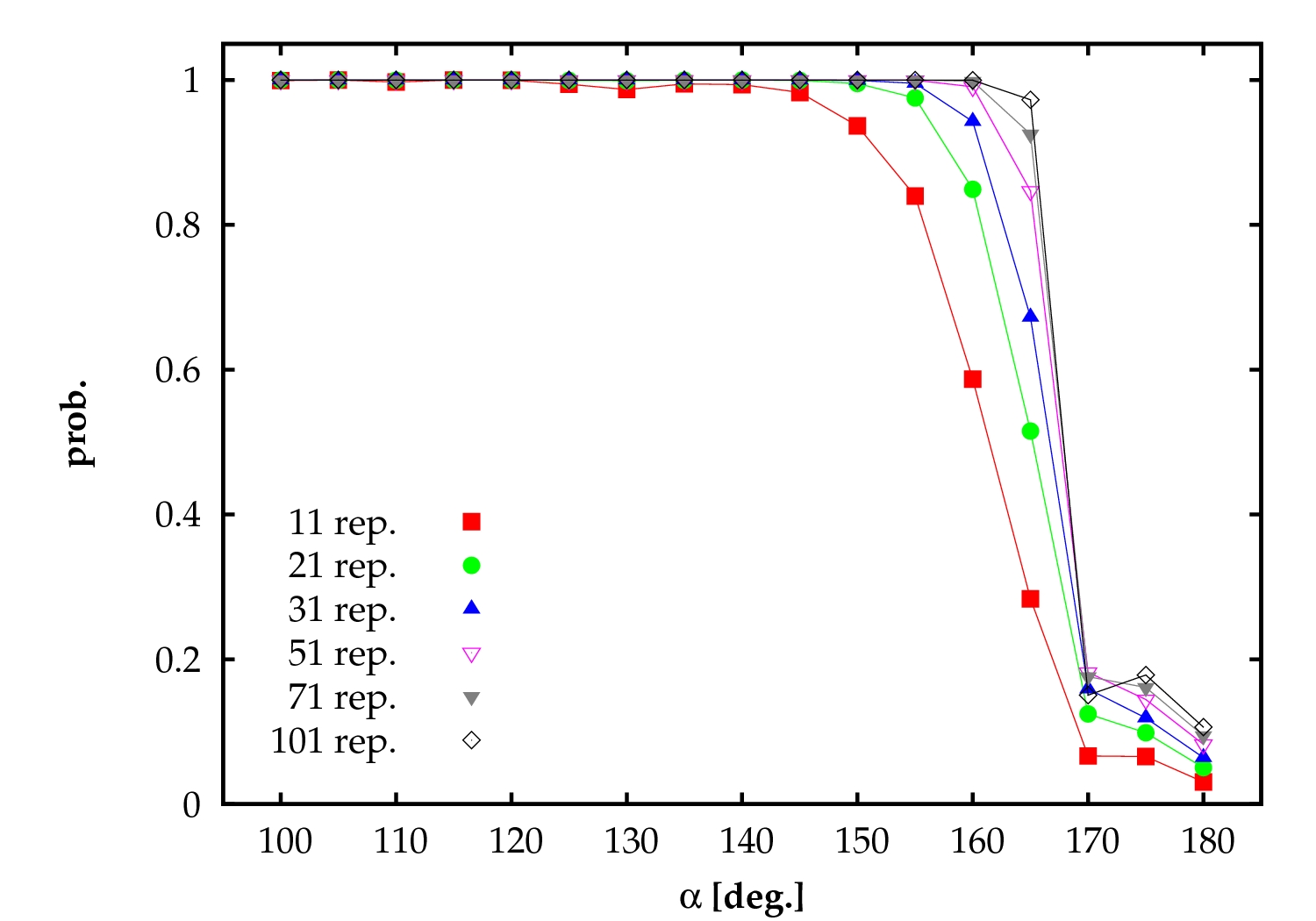}    
   a) HF guess 
 \end{center} 
\end{minipage}
\begin{minipage}{0.45\linewidth}
 \begin{center}
   \includegraphics[width=7.5cm]{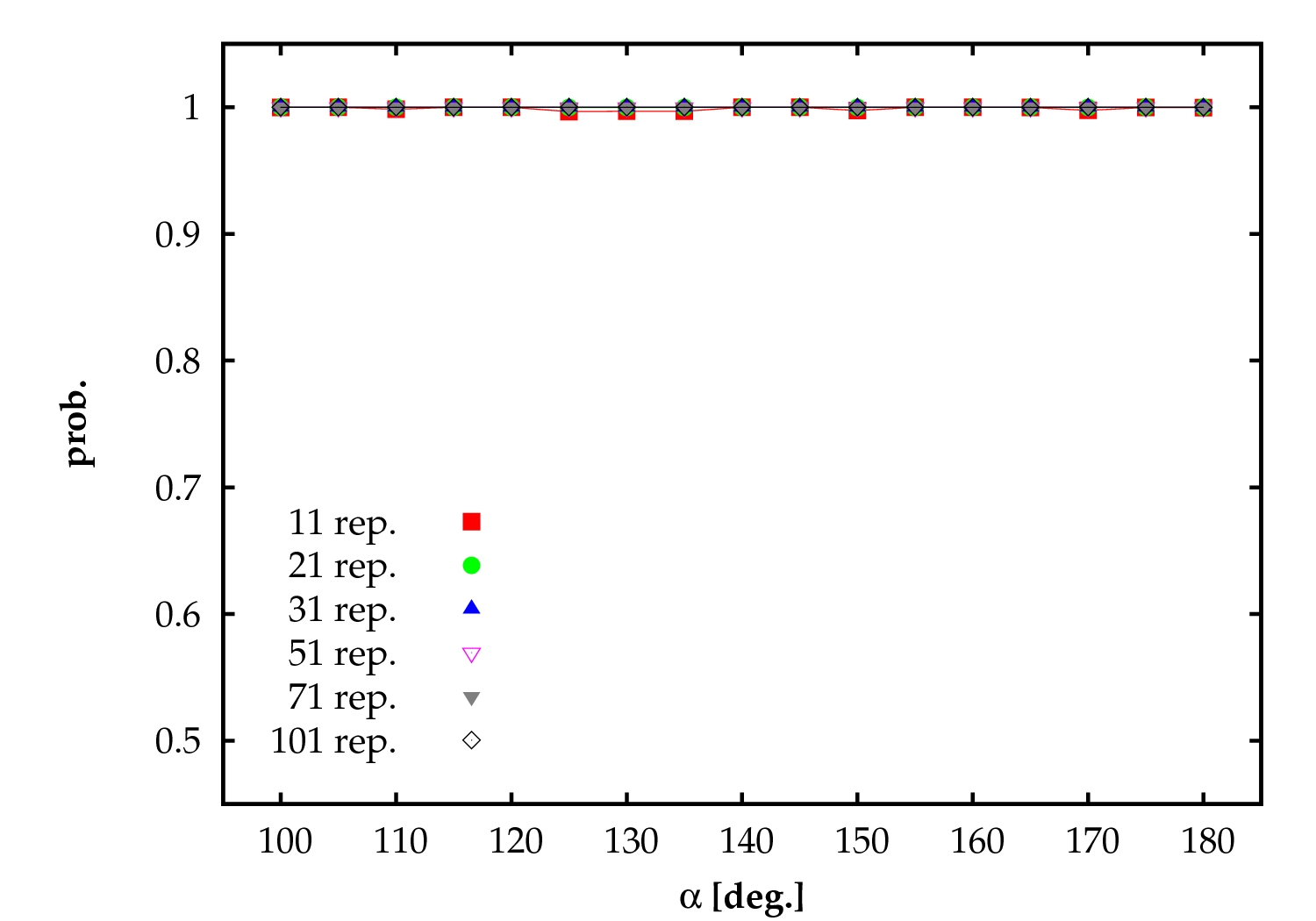}    
   b) CAS(2,2), tresh. 0.2 guess 
 \end{center}
\end{minipage}
\caption{Success probabilities of the \textbf{B} version of IPEA with HF and CAS(2,2), tresh. 0.2 initial guesses and different number of repetitions of individual bit measurements for the $\tilde{a}~^{1}A_{1}$ state, $\alpha$ denotes the H-C-H angle.}
\label{bend_alg1}
\end{figure}

Results for the H-C-H angle bending are summarized in Figures \ref{bend_alg0} and \ref{bend_alg1}. Simulations concerning this process involve only the $\tilde{a}~^{1}A_{1}$ state as this state exhibits a strong mul\-ti\-re\-fe\-rence character when going to linear geometries. In this case, no swapping of molecular orbitals occurs during the process and the complete active space CAS(2,2) was always constructed from $3a_{1}$  (HOMO) and $1b_{1}$ (LUMO) molecular orbitals. Moreover, due to the different symmetry of these orbitals, only two configurations contribute to CAS(2,2) wave function: doubly occupied HOMO [configuration (\ref{conf2})] and doubly occupied LUMO [configuration (\ref{conf3})]. Both of these configurations have for all values of $\alpha$ (H-C-H angle) absolute values of amplitudes higher than 0.2. 

Figure \ref{bend_alg0} presents the results of the \textbf{A} version of IPEA. Overlap and scaled overlap of the initial HF guess wave function and the exact FCI wave function is again shown as well as the success probabilities for HF and CAS(2,2), tresh. 0.2 guesses and dotted line bounding the safe region. Performance of the \textbf{B} version of IPEA with HF and CAS(2,2), tresh. 0.2 initial guesses is illustrated in Figure \ref{bend_alg1}.

\section{Discussion}
\subsection*{IPEA - A version}
Results of the simulations with \textbf{A} version of IPEA numerically confirm that success probabilities always lie in the interval $\big| \langle \psi_{\rm{init}} | \psi_{\rm{exact}} \rangle \big|^{2} \cdot \big( 0.81, 1 \big>$, depending on the value of the remainder $\delta$ (\ref{reminder}). This algorithm can be safely used when the resulting success probability is higher than 0.5 (as it can then be amplified by repeating the whole process). We would like to note that when studying the applicability of this algorithm, one must monitor the scaled overlap between the initial guess and the exact wave function. 

Success probability higher than 0.5 is securely fulfilled with the HF initial guess at the equilibrium geometry for all four simulated states. When going to more stretched C-H bonds or linear geometry, RHF initial guess fails. CAS(2,2) initial guess improves the success probability in case of $\tilde{a}~^{1}A_{1}$ and $\tilde{c}~^{1}A_{1}$ states near the equilibrium geometry but in the region of more stretched C-H bonds it also fails. In this region, CAS(4,4) initial guesses must be used. For $\tilde{a}~^{1}A_{1}$ state, CAS(4,4), tresh. 0.2 guess is sufficient but for $\tilde{c}~^{1}A_{1}$, even the CAS(4,4) guess fails for few lengths of C-H bonds: $r/r_{0} = 2.2, 2.8 - 3.0$. For these points of the potential energy surface, where the overlap between the initial guess and the exact wave functions is not high enough, bigger active space should probably be used.

The situation is more difficult for the states of $B_{1}$ symmetry ($\tilde{X}~^{3}B_{1}$,  $\tilde{b}~^{1}B_{1}$) when the C-H bonds are stretched. Here even CAS(4,4) initial guess fails and bigger active space - CAS(4,5) - must be used for initial guess state calculations. In STO-3G basis set, this bigger complete active space contains five from the total number of seven molecular orbitals and represents therefore nearly the whole space. For this reason, we performed the classical FCI calculations with the cc-pVDZ basis set (\textit{1s} orbital on the carbon atom was kept frozen to reduce computational demands), where the total number of molecular orbitals is 24 (much more than the number of molecular orbitals in the complete active space), and verified that the overlap between the CAS(4,5) and the exact wave function is sufficiently high, essentially the same as in STO-3G basis set. Apart from the active space size, initial guess states always contained at most 12 configurations, but usually 8 or even less for nearly dissociated molecule. This observation is in agreement with the results of \cite{wang_2008}, where few configuration state functions added to the initial guess improved the success probability dramatically.

High success probabilities (over 0.8) can on the other hand be obtained with CAS(2,2), tresh. 0.2 initial guess for $\tilde{a}~^{1}A_{1}$ state during H-C-H angle bending. Initial guess states for this process correspond to only two configurations and are thus very easy to prepare (e.g. according to \cite{ortiz_2001}).

We would like to note one more thing about the \textbf{A} version of IPEA. This version can be in principle \textit{effectively} used for ground state energy calculations even when the success probability is lower than 0.5. In fact, since it always converges to one of the eigenstates, it can be used in all situations when the overlap between the initial guess state and the exact wave function is not exponentially small (with respect to size of the system). In those cases, one will repeat the procedure (a polynomial number of times, depending on the value of the aforementioned overlap) and look for the lowest energy. The only thing one must pay attention to is to set the limits $E_{\mathrm{max}}$ and $E_{\mathrm{min}}$ so that all the eigenvalues of the Hamiltonian lie between them. Then only it is sure that the lowest energy is not artificial and really corresponds to the lowest eigenvalue.

\subsection*{IPEA - B version}
This version of IPEA is characteristic by repeated initial state preparation in each iteration and has the disadvantage that no collapsing of the system and improving the overlap between the actual state of the quantum register and the exact wave function occurs. The situation is however not so bad because one does not ``fight" against the overlap between the initial guess and unwanted eigenfunctions at every iteration. This would happen only if all binary digits of the phase were opposite to binary digits of the phases of all other eigenstates, which is clearly not possible. Our simulations proved that a relatively small number of repetitions ($\approx 51$) at each iteration step give nearly unity success probability when modest-size initial guesses are used.

The results with the RHF guesses (Figures \ref{singlet_a1_lower_rep_hf} and \ref{bend_alg1}a) nicely show their limits (for $\tilde{a}~^{1}A_{1}$ state). These are: $r/r_{0} = 2.3$ for C-H bond stretching and 170$^{\circ}$ for H-C-H angle bending. We have chosen the ``best" initial guess states in terms of price performance ratio for each of the four electronic states: $\tilde{a}~^{1}A_{1}$ (C-H bond stretching), $\tilde{c}~^{1}A_{1}$: CAS(4,4), tresh. 0.2; $\tilde{X}~^{3}B_{1}$, $\tilde{b}~^{1}B_{1}$: CAS(4,5), tresh 0.2; $\tilde{a}~^{1}A_{1}$ (H-C-H angle bending): CAS(2,2), tresh. 0.2. They were chosen to contain the minimum number of configurations, yet give high enough success probabilities. These initial guesses performed very well and the only exception are a few points for $\tilde{c}~^{1}A_{1}$ state, where a bigger active space is desirable.

We have not simulated the decoherence phenomena and when taking it into account the situation will surely change. Quantum error correction \cite{gaitan_book} would probably be needed for the \textbf{A} version, which would increase the number of required qubits as well as the complexity of the quantum circuit, while \textbf{B} version should be more robust.

\section{Conclusions}
We have developed programs (C++) for simulation of a quantum computer and successfully performed QFCI energy calculations of ground as well as excited states of CH$_{2}$ molecule that exhibits multireference character. We have demonstrated that energies at equilibrium geometry are accessible with RHF initial guesses, which are easy to prepare. CASCI initial guess states with small complete active spaces composed of relatively few configurations ($\approx 10$) are sufficient even for a nearly dissociated molecule to achieve the probability amplification regime of the IPEA algorithm. Its version \textbf{B} with repeated initial state preparation seems to be a better candidate for the first real larger-scale QFCI calculations than the \textbf{A} version, since it does not require a long coherence time. 

\section*{Acknowledgment}
This work has been supported by the Grant Agency of the Czech Republic - GA\v{C}R (203/08/0626) and the Grant Agency of the Charles University - GAUK (114310).

\bibliography{kvantove_pocitace,clanek_ch2,gelfand,cc}

\clearpage

\begin{figure}[!h]
 \begin{center}  
 \mbox{
   \Qcircuit @C=1em @R=1em {
     \lstick{\ket{0}} & \gate{H} & \ctrl{1} & \gate{R_{z}(\omega_{k})} & \gate{H} & \meter & \cw & \phi_{k} \\
     \lstick{\ket{u}} & {/} \qw & \gate{U^{2^{k-1}}} & {/} \qw & \qw
   }
 }
 \end{center}
 \caption{The $k$-th iteration of the iterative phase estimation algorithm (IPEA). The feedback angle
depends on the previously measured bits. Note that $k$ is iterated backwards from $m$ to 1.}
 \label{ipea_iteration} 
\end{figure}
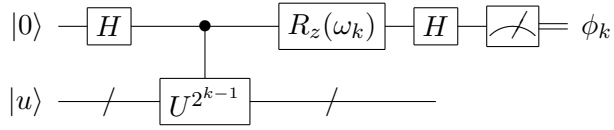

\begin{figure}[!h]
 \begin{minipage}{\linewidth}
 \centering
 \mbox{
   \Qcircuit @C=0.7em @R=1em {
     & \lstick{\ket{0}} & \gate{H} & \ctrl{1} & \gate{R_{z}(\omega_{m})} & \gate{H} & \meter & \cw & ~x_{m} & & & & &  \lstick{\ket{0}} & \gate{H} & \ctrl{1} & \gate{R_{z}(\omega_{m-1})} & \gate{H} & \meter & \cw & & x_{m-1} \\
     & \gate{\rm{ISP}} & {/} \qw & \gate{U^{2^{m-1}}} & \qw & \qw & \qw & {/} \qw & \qw & \qw & \qw & \qw & \qw & \qw & \qw & \gate{U^{2^{m-2}}} & {/} \qw & \qw
   }
 }
 \vskip 0.3cm  
 a) Maintaining the second part of the quantum register during all iterations.
 \end{minipage}

 \vskip 0.5cm

 \begin{minipage}{\linewidth}
 \centering
 \mbox{
   \Qcircuit @C=0.7em @R=1em {
     & \lstick{\ket{0}} & \gate{H} & \ctrl{1} & \gate{R_{z}(\omega_{m})} & \gate{H} & \meter & \cw & ~x_{m} & & & &  \lstick{\ket{0}} & \gate{H} & \ctrl{1} & \gate{R_{z}(\omega_{m-1})} & \gate{H} & \meter & \cw & & x_{m-1} \\
     & \gate{\rm{ISP}} & {/} \qw & \gate{U^{2^{m-1}}} & {/} \qw & \qw & & & & & & & \gate{\rm{ISP}} & {/} \qw & \gate{U^{2^{m-2}}} & {/} \qw & \qw
   }  
 }  
 \vskip 0.3cm    
 b) Initial state preparation per iteration.
 \end{minipage}  
 \caption{Comparison of the two versions of IPEA. Oracle \fbox{ISP} prepares the initial state (\textit{Initial State Preparation}).}
 \label{ipea_comparison}
\end{figure}

\end{document}